\documentclass[preprint2,longabstract]{aastex}

\newcommand{\myemail}{quanz@astro.phys.ethz.ch}

\shorttitle{A large, massive, rotating disk around an isolated young stellar object}
\shortauthors{Quanz et al.}

\begin{document}


\title{A large, massive, rotating disk around an isolated young stellar object}


\author{Sascha P. Quanz}
\affil{Institute for Astronomy, ETH Zurich, Wolfgang-Pauli-Strasse 27, 8093 Zurich, Switzerland\\Max Planck Institute for Astronomy, K\"onigstuhl 17, 69117 Heidelberg, Germany}    
\email{\myemail}
\author{Henrik Beuther}
\affil{Max Planck Institute for Astronomy, K\"onigstuhl 17, 69117 Heidelberg, Germany}  
\author{J\"urgen Steinacker}
\affil{LERMA, Observatoire de Paris, 61 Av. de l'Observatoire, 75 014 Paris, France\\Max Planck Institute for Astronomy, K\"onigstuhl 17, 69117 Heidelberg, Germany}
\author{Hendrik Linz,}
\affil{Max Planck Institute for Astronomy, K\"onigstuhl 17, 69117 Heidelberg, Germany}  
\author{Stephan M. Birkmann}
\affil{ESA/ESTEC, Keplerlaan 1, Postbus 299, 2200 AG Noordwijk, The Netherlands}
\author{Oliver Krause, Thomas Henning}
\affil{Max Planck Institute for Astronomy, K\"onigstuhl 17, 69117 Heidelberg, Germany}  
\and
\author{Qizhou Zhang}
\affil{Harvard-Smithsonian Center for Astrophysics, Cambridge, Massachusetts, USA}
\altaffiltext{1}{Based on observations made at the Calar Alto Observatory.
This work is based in part on data collected at Subaru Telescope, which is operated by the National Astronomical Observatory of Japan, and on observations made with the \emph{ Spitzer Space Telescope}, which is operated by the Jet Propulsion Laboratory, California Institute of Technology under a contract with NASA.}


\begin{abstract}
We present multi-wavelengths observations and a radiative transfer model of a newly discovered massive circumstellar disk of gas and dust which is one of the largest disks known today. Seen almost edge-on, the disk is resolved in high-resolution near-infrared (NIR) images and appears as a dark lane of high opacity intersecting a bipolar reflection nebula. Based on molecular line observations we estimate the distance to the object to be 3.5 kpc.  This leads to a size for the dark lane of $\sim$10500 AU but due to shadowing effects the true disk size could be smaller. In \emph{Spitzer/IRAC} 3.6 $\mu$m images the elongated shape of the bipolar reflection nebula is still preserved and the bulk of the flux seems to come from disk regions that can be detected due to the slight inclination of the disk. At longer \emph{IRAC} wavelengths, the flux is mainly coming from the central regions penetrating directly through the dust lane. Interferometric observations of the dust continuum emission at millimeter wavelengths with the SMA confirm this finding as the peak of the unresolved mm-emission coincides perfectly with the peak of the  \emph{Spitzer/IRAC} 5.8 $\mu$m flux and the center of the dark lane seen in the NIR images. Simultaneously acquired CO data reveal a molecular outflow along the northern part of the reflection nebula which seems to be the outflow cavity.  An elongated gaseous disk component is also detected and shows signs of rotation. The emission is perpendicular to the molecular outflow and thus parallel to but even more extended than the dark lane in the NIR images. Based on the dust continuum and the CO observations we estimate a disk mass of up to a few solar masses depending on the underlying assumptions. Whether the disk-like structure is an actual accretion disk or
rather a larger-scale flattened envelope or pseudodisk is difficult to
discriminate with the current dataset.
The existence of HCO$^+$/H$^{13}$CO$^+$ emission proofs the presence of dense gas in the disk and the molecules' abundances are similar to those found in other circumstellar disks. We furthermore detected C$_2$H towards the objects and discuss this finding in the context of star formation. 
Finally, we have performed radiative transfer modeling of the K band scattered 
light image varying a disk plus outflow 2D density profile and the 
stellar properties. The model approximately reproduces extent and
location of the dark lane, and the basic appearance of the outflow.
We discuss our findings in the context of circumstellar disks across all mass regimes and conclude that our discovery is an ideal laboratory to study the early phases in the evolution of massive circumstellar disks surrounding young stellar objects.

\end{abstract}





\section{Introduction}
In recent years, significant progress has been made in our understanding of Young Stellar Objects (YSOs) and the formation of stars. Still, there remain unanswered questions and problems that have to be solved observationally and theoretically. In particular, while the formation of low-mass and intermediate mass stars 
by accretion of material via a circumstellar disk is well established \citep[e.g.,][]{watson2007}, for 
massive stars (M $\ge$ 8 M$_\sun$) not only the theory of the star formation process 
\citep[e.g.,][]{stahler2000} but also the observations are challenging \citep[e.g.,][]{beuther2007,zinnecker2007} and yield
partly contradicting results \citep[e.g.,][]{chini2004,sako2005,steinacker2006,nurnberger2007}. Conceptually the problem arises from the fact that at least in the spherically  symmetric case the strong radiation pressure of a hydrogen burning young massive star with a mass  $\ge$ 8 M$_\sun$ would prevent any further gas infall and would thus halt the formation of a more massive object. 
Two competing theories overcoming this problem were intensively discussed in the recent past. They explain the formation of massive stars either by the so-called "Competitive Accretion and Merging Model" \citep{bonnell2006} or by fragmentation during a large scale collapse in turbulent cores \citep{mckee2003}. The latter model resembles the formation process of low-mass objects and consequently relies on circumstellar disks around young massive stars to build up the final objects \citep[see,	 also,][]{krumholz2009}.

Today, there is ample evidence for the existence of "disk-like" structures around massive young stellar objects 
with masses/sizes being higher/bigger compared to typical low-mass circumstellar disks 
\citep[e.g.,][]{cesaroni2007,schreyer2006,chini2004,beltran2004,patel2005,jiang2005,jiang2008}. While for early B-stars these structures appear to be indeed rather disk-like, for O-stars they appear to have more of a toroidal shape. In most cases large, flattened structures were detected using high spatial resolution molecular line data in the (sub-)mm regime. Typically, at shorter wavelengths, the dense envelope surrounding the massive protostar causes too much extinction and prevents a direct detection of the source in the mid- or near-infrared. Indications for rotation in the flattened gaseous structures supported the hypothesis that these objects are massive circumstellar disks. However, the question whether these objects are indeed the high-mass counterpart to the well-established low-mass protoplanetary accretion disks directly feeding the central source or whether they are (remnants of) the rotating and infalling molecular envelopes \citep[e.g.,][]{beuther2009} or "pseudodisks"\footnote{A "pseudodisk" is a flattened disequilibrium structure created during the collapse of a magnetized cloud core by strong magnetic pinching forces deflecting the infalling gas toward the equatorial plane \citep{gallishu1993}.} \citep{gallishu1993,allen2003} is unclear. In particular, it appears likely that at least some of the observed massive gaseous "toroids" around stars of 10-20 M$_\sun$ are unstable and have comparatively short lifetimes \citep{cesaroni2007} while low-mass disks typically persist for several Myr. 

In addition to (sub-)mm data, only for very few massive objects mid- or near-infrared (MIR / NIR) data exist that support the existence of an extended dusty disk component. \citet{chini2004} presented an NIR image of an assumed accretion disk surrounding a presumably massive protostar in M17, but their interpretation was challenged by \citet{sako2005}. A detailed radiative transfer modeling of the $K$-band image of this source was performed in Steinacker et al. (2006) showing that the underlying structure has a disk shape with a mass around a few M$_\odot$.
Using NIR polarimetric images \citet{jiang2005,jiang2008} found elongated polarized signatures around massive young stellar objects suggesting the presence of polarization disks. \citet{sridharan2005} claimed the detection of a silhouette edge-on disk around IRAS 20126+4104 based on NIR images in the $K$, $L'$, and $M'$ filters. More recently, however, \citet{debuizer2007} derived from MIR images at 12.5 and 18.3\,$\mu$m a much more complex structure of the inner regions of the same source and suggested that the data could also be interpreted having a tight young stellar cluster in the center. Based on MIR images, \citet{chini2006} reported the detection of a remnant disk around a young massive star, and \citet{nielbock2007} reported on evidence for a disk candidate 
surrounding a hypercompact HII region in M17 based on NIR data. These authors also provided a radiative transfer model for the disk seen in the $K_s$-band image.
Finally, very recently \citet{okamoto2009} published the detection of a flared circumstellar disk around the 10 M$_\sun$ Herbig Be star HD200776 also using MIR imaging data. 

Here, we report the direct detection of a large and massive circumstellar disk of dust and gas surrounding a young stellar source. 
The object is named CAHA J23056+6016 as it was first identified in NIR images taken at the Calar Alto observatory (CAHA). Currently, we can not say whether the disk is an actual accretion disk or rather a "pseudodisk" or a flattened envelope and we use the more general term "circumstellar disk" or simply "disk" throughout the paper to avoid any possible misinterpretation. The morphology shows a bipolar nebula intersected by a dark lane. Multi-wavelengths data including NIR/MIR images as well as dust continuum and molecular line data are used to constrain the physical properties of this object. We derive a distance of 3.5 kpc to the source, find a radius for the disk of several thousand AU and a mass, depending on the underlying assumptions, of up to several solar masses. The $K-$band and the mm image are analyzed with a disk/outflow model using radiative transfer calculations to derive scattered light
images.

In section 2 we describe the observations and data reduction and present the results of the data in section 3. Results from the radiative transfer model are presented in section 4. 
A discussion about the nature and properties of CAHA J23056+6016 in the context of star formation is given in section 5. Finally, we summarize our findings and conclude in section 6. 

\section{Observations and data reduction}
\subsection{{\sc Omega2000} NIR wide-field images} 
CAHA J23056+6016 was detected in a deep NIR survey in October 2003 with the wide-field NIR 
camera {\sc Omega2000} on the 3.5 meter telescope in Calar Alto (Spain)\footnote{http://w3.caha.es/CAHA/Instruments/O2000/index2.html}. 
The camera is equipped with a 2048$\times$2048 pixel HAWAII-2 detector, has a pixel scale of 0.45$''$/pixel and an effective field-of-view of $\sim$15$'$. Images were taken in the three NIR filters $J$, $H$, and $K_s$ with central wavelengths at 1.21, 1.65, and 2.15\,$\mu$m, respectively. The observations followed a pre-defined dither pattern of 20 positions around a defined image center. On each position 20 frames with an integration time of 3 seconds were taken, summing up to 1 minute integration time per dither position. After co-addition of all images the total integration time amounted to 20 minutes per filter. 

Flat fielding, bad pixel correction and co-addition of the images were done with standard {\sc IRAF}\footnote{http://iraf.noao.edu/} routines. Furthermore we used the {\sc drizzle} routine to reduce the pixel scale of the final images to half of its original value, i.e., to 0.225$''$/pixel. Astrometry was applied by comparing the positions of detected 2MASS point sources to those listed in the
2MASS Point Source Catalog (PSC) \citep{cutri2003}. On average the precision of the astrometry is better than 0.25$''$ with respect to the known 2MASS positions.

The photometric calibration of CAHA J23056+6016 and its nebula (see section 3.1) was done using six 2MASS sources  in the immediate vicinity of CAHA J23056+6016 as reference stars (for the J filter image seven reference sources were available). The IDL routine {\sc atv.pro} was used to carry out aperture photometry and derive count rates of the sources in the {\sc Omega2000} images. The 2MASS magnitudes of the objects were then converted to mJy using the Magnitude-to-Flux Density converter provided by the \emph{Spitzer} Science Center\footnote{http://ssc.spitzer.caltech.edu/tools/magtojy/}. Plotting flux density vs. count rate for all reference object and fitting an error-weighted line though the data allowed us to derive the relation between counts/pixel and mJy/per pixel. For each filter this relation was used to calibrate our {\sc Omega2000} images. As a cross-check we performed aperture photometry on the calibrated images to re-derive the flux density for the reference sources. For each filter the results were within the error bars from the 2MASS photometry giving us confidence in our photometric calibration. 

\subsection{{\sc Subaru IRCS} high-resolution NIR images} 
In June 2008 high-resolution $H$- and $K'$-band images were obtained with the Infrared Camera and Spectrograph (IRCS) at the Subaru telescope. IRCS is equipped with a 1024$\times$1024 pixel Alladin3 detector providing a pixel scale of $\sim$0.027 mas/pixel. In each filter 7 sets of 9 dither positions were observed. On each dither position 5 exposures of 5 seconds were taken resulting in a total integration time of 1575 sec ($=$26.25 min) per filter. The data processing and photometric calibration was done in the same manner as we did for the wide-field images. We reached a 5-$\sigma$ limiting magnitude of $\sim$22.7 mag in $H$ and $\sim$23.1 mag in $K$ in the 2MASS photometric system.

\subsection{\emph{Spitzer IRAC} images} 
\emph{Spitzer IRAC} data were downloaded from the \emph{Spitzer} data archive (AOR-Key: 18917120). The data were taken in February 2007 and serendipitously covered CAHA J23056+6016. As the observing parameters were not optimized for CAHA J23056+6016 but rather for bright, nearby HII regions, the integration time per pixel was only 1.2 seconds. Hence, for CAHA J23056+6016 the data are of modest signal-to-noise and only data from channel 1, 2 and 3 (at 3.6, 4.5 and 5.8\,$\mu$m, respectively) are used in our analyses. We used MOPEX\footnote{http://ssc.spitzer.caltech.edu/postbcd/download-mopex.html} to carry out the basic data reduction steps (outlier rejection, mosaicking) and create the final images. In this process, the pixel scale was changed to 0.6$''$/pixel in all channels instead of the default value of 1.2$''$/pixel. To convert the flux densities from the initially given unit of MJy/sr to $\mu$Jy/pixel (similar to the NIR data) we adopted 
the value provided by the \emph{Spitzer} Science Center\footnote{http://ssc.spitzer.caltech.edu/archanaly/quickphot.html} to the new pixel scale:
\begin{equation}
1\;{\rm MJy/sr} \approx 8.46161\;\mu{\rm Jy/pixel}
\end{equation}

\subsection{Millimeter wavelength observations}
\subsubsection{SMT} 
Following the initial discovery of CAHA J23056+6016 in 2003 we obtained first radio data at the 10-m Sub-Millimeter Telescope (SMT) on Mount Graham in December 2004. 
Molecular line measurements of CO(3--2) (beam size $\sim$22$''$) were obtained at 345 GHz. We used the facility 
receivers and reduced all spectra with the CLASS software\footnote{http://www.iram.fr/IRAMFR/GILDAS}. The 
atmospheric transmission was $\tau_{225\,{\rm GHz}}\approx 0.2$ and W75N was used as 
calibration source. The estimated pointing error is 3$''$ and the 
radiometric accuracy is 15\%.

\subsubsection{IRAM 30\,m telescope}\label{iram_obs}
In June 2006 we were granted DDT time at the IRAM 30-m telescope on Pico Veleta (Spain) to observe CAHA J23056+6016 in mm dust continuum with MAMBO1 in on-off mode. The MAMBO1 bolometer works at 1.2 mm, has 37 pixels and a pixel spacing of 20$''$. The half-power-beam-width (HPBW) is $\sim$10.5$''$. The total on-source integration time amounted to 40 minutes. Cep A and NGC7538 served as calibration sources. The data reduction was done using standard routines from the MOPSIC software package.

Furthermore, the IRAM 30\,m telescope was used to obtain spectra of CAHA J23056+6016 in the molecular lines 
HCO$^+$(1--0) at 89.189 GHz and its isotopologue H$^{13}$CO$^+$(1--0) at 86.754 GHz. These 
observations took place between August 20 - 24, 2009. The EMIR receiver E90 was utilized in combination 
with the VESPA backend.  Position switching was employed with on-source times of 120 s and 840 s for 
HCO$^+$(1--0) and H$^{13}$CO$^+$(1--0), respectively. The spectral resolution was 40 kHz per channel, 
which corresponds to roughly 0.135 km/s. The data were first calibrated to the antenna temperature scale 
using the common chopper-wheel method and then converted into main beam brightness temperatures using the forward and main beam efficiencies provided by the observatory\footnote{http://www.iram.es/IRAMES/mainWiki/Iram30mEfficiencies}. The achieved rms noise levels (in T$_{\rm mb}$) are 0.043 K for HCO$^+$(1--0) and 0.018 K for H$^{13}$CO$^+$(1--0), respectively.

The C$_2$H(1--0) transitions were observed in July 2009 at the IRAM 30\,m 
telescope with the new receiver system EMIR. All hyperfine structure lines 
near 87.4\,GHz were observed \citep{bel1998} with a spectral resolution 
of $\sim$0.13\,km\,s$^{-1}$. System temperatures at 3\,mm wavelengths 
were 88\,K.  The 
1$\sigma$ rms of the final smoothed spectrum with 0.27\,km\,s${^-1}$ 
spectral resolution is 25\,mK.

\subsubsection{SMA}
CAHA J23056+6016 was observed with the Submillimeter Array (SMA)\footnote{The Submillimeter Array is a joint project between the Smithsonian  Astrophysical Observatory and the Academia Sinica Institute of Astronomy and Astrophysics, and is funded by the Smithsonian Institution and the Academia Sinica.}
on October 7th 2007 with 8 antennas at 1.3\,mm in the compact configuration 
with projected baselines up to 53\,k$\lambda$. The phase center was 
23$^h$05$^m$37.4$^s$ in RA, and $60^{\circ}15'45''.6$ in DEC (J2000.0) with a 
tuning frequency of 230.537970\,GHz in the upper sideband and chunk 14 
($v_{\rm{lsr}}=-49.6$\,km\,s$^{-1}$). The spectral resolution was 
0.406\,kHz/channel corresponding to $\sim 0.55$\,km\,s$^{-1}$ velocity 
resolution. The weather was intermediate with zenith opacities 
$\tau(225\rm{GHz})$ around 0.15 measured by the National Radio Astronomy 
Observatory (NRAO) tipping radiometer operated by the Caltech 
Submillimeter Observatory (CSO). Passband and flux calibration were 
derived from 3C111 and Ganymede observations, and the flux density scale 
is estimated to be accurate within 20\%. Phase and amplitude were 
calibrated with regularly interleaved observations of the quasar 0102+584. 
We applied different weightings for the continuum and line data resulting 
in synthesized beams of $4.3''\times 3.7''$ for the mm continuum emission 
and $3.8''\times 3.3''$ and $4.0''\times 3.5''$ for the C$^{18}$O(2--1) 
and $^{12}$CO(2--1) emission, respectively. The rms noise of the line and 
continuum images are 1.4\,mJy\,beam$^{-1}$ and 150\,mJy\,beam$^{-1}$ per 
0.6\,km\,s$^{-1}$ wide channel. The initial flagging and calibration was 
done with the IDL superset MIR originally developed for the Owens Valley 
Radio Observatory \citep{scoville1993} and adapted for the 
SMA\footnote{The MIR cookbook by Charlie Qi can be found at 
http://cfa-www.harvard.edu/$\sim$cqi/mircook.html.}. The imaging and data 
analysis were conducted in MIRIAD \citep{sault1995}.

\section{Results}
\subsection{NIR images}
\subsubsection{The large scale environment}\label{large_scale}
Figure~\ref{color-image-wide} shows a color composite of the {\sc Omega2000} NIR images with CAHA J23056+6016 lying at the image center (RA (J2000): 23h 05m 37.52s, DEC (J2000): +60$^\circ$ 15$'$ 45.8$''$). A bipolar, cone-shaped nebula and an intersecting dark lane are clearly visible. In the extended surroundings of CAHA J23056+6016 there is evidence for ongoing star formation: The large HII complex in the lower right corner of the image is IC 1470 (alias IRAS 23030+5958 or S156) with a distance of $\sim 3.6'$ from CAHA J23056+6016, and the bright star in the lower left corner is WRAM17 (distance $\sim 2.7'$) which is assumed to be the exciting source of BFS17, a compact HII region in its immediate vicinity \citep{russeil2007}. Furthermore, IRAS 23033+5951, another compact high-mass star-forming region, is lying $\sim 7.7'$ south of CAHA J23056+6016. 

\subsubsection{Source morphology - an almost edge-on disk}
Figure~\ref{nir-images} focuses directly on the detailed structure of CAHA J23056+6016 in the NIR. While the $J$ filter image was taken at Calar Alto, the $H$ and $K'$ filter images were obtained with Subaru and superior spatial resolution. The contours give an idea of the different observed flux levels and the achieved S/N in each filter. Assuming that the flux is mostly scattered light from a dusty surface (see below) the fact that the nebula appears brighter at longer NIR wavelengths is surprising as the scattering efficiency increases for shorter wavelengths for typical ISM dust grains. However, line-of-sight extinction towards the source causes higher absorption at short wavelengths and could explain the difference in the apparent brightness of the nebula.

The high-resolution images in the middle and right-hand panel in Figure~\ref{nir-images} clearly resolve the bipolar nebula. The nebula is rotated in the plane of the sky by roughly $\sim$10$^{\circ}$ from north to the west and stretches over $\sim$12$''$ in the north--south and $>$6$''$ in the east--west direction. The northern and southern cones of the nebula are asymmetric in size and brightness, with the northern part being more prominent in size and luminosity. Assuming an intrinsic symmetry of the structure this can be explained if the northern part is tilted towards the observer compared to the southern part (see also sections~\ref{irac_images} and \ref{outflow}) and if forward scattering from dust grains is the main source of brightness of the nebula in the NIR. Forward scattering and the suggested orientation would also explain the rather smooth brightness distribution and the position of the brightest regions in the nebula. For the scattering to work, an internal luminosity source must be present illuminating its dusty, cone shaped surroundings. The central source, however, is not detected in the NIR images directly but is presumably hidden behind the flattened structure of high extinction intersecting the northern and southern part of the nebula. As can be seen in Figure~\ref{nir-images} the degree of extinction of this intersecting lane depends on the wavelengths as would be expected if dust was the main source of extinction: Even if the $J$-band image has a lower S/N than the other images the gap between the northern part of the nebula and the southern part appears much wider. Going to longer wavelengths, the apparent size of the gap decreases in $H$, and with the chosen contour levels the gap is no longer apparent in the $K'$ image. However, as can be nicely seen in the $K'$ band, the contours suggest a layered structure of extinction having its maximum in the middle and then slowly decreasing towards the north and the south. Putting all this together, CAHA J23056+6016 shows a striking similarity to edge-on circumstellar disk systems found around young low-mass stars with a characteristic bipolar nebula and an intersecting dark dust lane \citep[e.g.,][]{padgett1999,brandner2000}. However, as mentioned above, due to the asymmetry and rotation of the nebula the disk of CAHA J23056+6016 appears to be slightly inclined and tilted. 

The apparent size of the dark lane at the base of the northern part of the nebula is $\sim$3$''$. This translates into $\sim$10500 AU for an assumed distance of 3.5 kpc (see section~\ref{distance_estimate} for details on the derivation of the distance). 
However, the dark lane is caused by the extinction of scattered light in the disk and what we observe could actually be not the disk itself but rather its shadow onto the surrounding nebula \citep{pontoppidan2005b}. Hence, in this case, we can not derive the disk size directly from the images and as described in section 4.1 the actual disk size may indeed be a bit smaller. 

The nebula in Figure~\ref{nir-images} has a physical size of $\sim42000\times21000$ AU or $\sim0.2\times0.1$ pc and appears to be more than one order of magnitude larger than those detected for edge-on systems in nearby star-forming regions \citep[e.g.,][]{padgett1999,brandner2000}.

\subsection{IRAC images - peaking through the dust lane}\label{irac_images}
Going to slightly longer wavelengths, Figure~\ref{irac-images} shows \emph{Spitzer/IRAC} images of CAHA J23056+6016. As mentioned in section 2.3. no optimized pointed observations were carried out but CAHA J23056+6016 was rather serendipitously imaged during a larger mapping program. Hence, the S/N is limited, especially in the images at 3.6 and 4.5 $\mu$m (see figure caption for details). 

At 3.6 $\mu$m the morphology of the emission still slightly resembles the overall morphology in the NIR. The emission is elongated in north-south direction and appears to be rotated slightly westwards. At this moment, we do not know whether the observed flux at this wavelength is mainly coming from PAH emission or whether it is caused by scattering. 
At 4.5 $\mu$m the emission becomes already more compact and the rotation of the elongated structure is gone. As mentioned above the emission in the third channel at 5.8 $\mu$m is rather intense and, similar to channel 2, the morphology does no longer resemble the NIR images. 
Two things are noteworthy regarding the IRAC images: (1) Comparing the peak of the emission in each channel one finds that it apparently changes its position. Coming from 3.6 $\mu$m the emission peak moves slightly southwards at 4.5 $\mu$m and then furthermore to the west at 5.8 $\mu$m. The overall shift between IRAC 1 and 3 is shown in Figure~\ref{nir-images} and appears to be $>0.8''$. This is significantly more than the expected astrometric accuracy of post-BCD\footnote{post Basic Calibrated Data} IRAC data which is better than $0.3''$ (see, \emph{Spitzer} webpage{\footnote{http://ssc.spitzer.caltech.edu/irac/products/overview.html}). Without emphasizing to much the derived numbers we believe that at least the trend is real. (2) Compared to the maximum emission in the NIR which was confined to the nebula, the maximum emission in the IRAC channel 2 and 3 is centered directly on the dark lane. 

Building upon the dust disk hypothesis from the NIR images these findings can be interpreted as follows. At longer IRAC wavelengths the disk is no longer opaque and one starts to detect the inner regions of the object potentially combining emission from the inner disk regions and the central source. 

The offset in the emission peak at 3.6 $\mu$m compared to 5.8 $\mu$m can be explained if one assumes, as laid out in section 3.1.2., that the disk is not seen edge-on but slightly inclined and rotated in the plane of the sky. The maximum flux at 3.6 $\mu$m would then not come from the central source because the opacity of the disk is still too high to directly "see" through the disk at this wavelength. Rather, the emission presumably traces the heated inner disk rim and those parts of the disk surface that are tilted towards the observer and are subject to less extinction. In Figure~\ref{nir-images} it can be seen that the emission peak at 3.6 $\mu$m is located at a position of higher luminosity and thus less extinction than the emission peak at 5.8 $\mu$m. Figure~\ref{color-image-zoom} shows a color-composite of $H$, $K'$ and IRAC 4.5 $\mu$m where the different regions of maximum emission become nicely evident.

\subsection{SMT, IRAM and SMA (sub-)mm data}
\subsubsection{Distance estimate}\label{distance_estimate}
To derive further information about CAHA J23056+6016 and to probe the properties of dust and gas in its vicinity, additional observations at even longer wavelengths were carried. In Figure~\ref{smt-spectrum} we show the result for the CO(3--2) molecular line observations at the SMT with the 345 GHz receiver. Several lines are detected suggesting the existence of multiple gaseous components towards the direction of CAHA J23056+6016. The strongest emission line is seen at a velocity of rest of -49.3 km/s and a slightly weaker line at -51.3 km/s (Table~\ref{lines}). The velocity of the second peak compares very well to the velocity we find for C$^{18}$O emission in our SMA data (see below). Thus, we believe that CAHA J23056+6016 is associated with emission at this velocity of rest and we assign a value of -51.3$\pm$0.2 km/s to the object.
Using the galactic rotation model of \citet{brand1986}\footnote{The model assumes a distance from the sun to the galactic center of 8.5 kpc and a velocity of the sun of 220 km/s.}  one finds a corresponding kinematic distance to the source of $\sim$5.1 kpc. The detected velocity is fairly consistent with those of the nearby HII regions mentioned in section 3.1.1: For IC 1470 and IRAS 23033+5951, \citet{bronfman1996} found velocities of -53.1 and -52.1 km/s, respectively. \citet{wouterloot1989} found for the later region a velocity of -51.8 km/s. Even if both HII regions lie $\sim$3.6$'$ and $\sim$7.7$'$ away from CAHA J23056+6016 one can assume that all belong to the same molecular cloud complex. In particular, \citet{russeil2007} concluded that their star-forming complex named "110.1+00" (based on its galactic coordinates) has a velocity of -52.0 km/s and is comprised of several HII regions including IC 1470 and also BFS 17, the third HII region close to CAHA J23056+6016 as mentioned in section 3.1.1. 
However, these authors also revise the distance to individual HII regions based on the spectral type and magnitudes of the exciting sources of the regions. For IC 1470 and BFS 17, they derive distances of 2.87$\pm$0.75 kpc and 3.66$\pm$1.7 kpc, respectively, which is significantly less than the 5.1 kpc derived above from the galactic rotation model. Furthermore, for IRAS 23033+5951 \citet{reid2008} give a distance of 3.5 kpc which they assume for the whole Cepheus star-forming complex. 
To be in agreement with the most recent estimates for the other regions, we adopt a distance of 3.5 kpc for CAHA J23056+6016. 

\subsubsection{Disk mass estimate from dust continuum}\label{disk_mass_section}
Additional mm-observations were carried out with MAMBO1 at the IRAM 30\,m telescope and finally with the SMA. The MAMBO1 observations at 1.2 mm continuum yielded a flux of 12.9$\pm$1.0 mJy with a beam FWHM of $\sim$10.5$''$. Hence, the beam covered not only the dusty disk in the center but also the nebula. During interferometric follow-up observations with the SMA with a synthesized beam of $\sim$4$''$ we measured a flux of 11.1 mJy/beam at 1.3 mm. The almost identical flux levels of the single dish and the interferometric observations show, that by far the bulk of the dust emission is coming from the central regions of the object, i.e. the disk. Since we did not resolve CAHA J23056+6016 with our SMA observations this emission comes from within the inner 4$''$. Figure~\ref{mm-image} shows an overlay of the SMA mm-data on the $H$-band Subaru image. The peak of the mm dust emission coincides very well with the IRAC 5.8 $\mu$m flux and the dark lane seen in the NIR images (Figure~\ref{nir-images}). Hence, the mm data confirm that the dark lane is indeed a dusty disk. Since the disk is not resolved in our SMA data, we can only give an upper limit on its size based on the mm observations. For a distance of 3.5 kpc we find the radius of the dust disk to be $\le 7000$ AU.

Assuming on optically thin configuration, we obtain for the dust mass 
\begin{equation}
M_{dust}=\frac{F_{1.3\,mm}\cdot d^2}{\kappa_{1.3\,mm}\cdot B_{1.3\,mm}(T_{dust})}
\end{equation}
where $F$ is the integrated flux, $d$ is the distance, $\kappa_{1.3\,mm}$ is the dust opacity in cm$^2$/g at 1.3 mm and $ B_{1.3\,mm}(T_{dust})$ is the Planck function at 1.3 mm for an assumed dust temperature $T_{dust}$. 
Since we do not know the temperature of the dust disk, nor the correct opacities for the grains, we estimate the total disk mass for a set of different parameters in Table~\ref{disk_mass}. Here, we always assume a flux of 11.1 mJy/beam and a distance of 3.5 kpc. The two values we use for the opacity correspond to that for dust grains with ice layers \citep{ossenkopf1994} and that assumed for circumstellar disks around low-stars, where in general some dust processing is assumed to have happend \citep{andrewswilliams2005}. Furthermore, to derive the total disk mass (gas and dust) from the dust mass, we multiply the values computed with equation (2) with the classical value for the gas-to-dust ratio of $M_{gas}/M_{dust}=100$, although a factor of $\approx$190 has been derived as median gas-to-dust ratio of nearby galaxies by \citet{draine2007}. The estimated range of disk masses reaches from $M_{disk}=0.4\,M_\sun$ to $M_{disk}=2.7\,M_\sun$, reflecting the uncertainties for mass estimates if dust opacities and temperatures are not well constrained. 
 
\subsubsection{Molecular outflow and hints for gas rotation}\label{outflow}
In addition to the mm continuum data, CO(2--1) molecular line observations were carried out at the SMA simultaneously. In Figure~\ref{mm-image} the green contours show the integrated CO(2--1) emission between -58 and -54 km/s. This blue-shifted elongated emission overlays nicely with the northern part of the nebula suggesting that the latter one is a cavity created by the molecular outflow. The fact that the line emission is blue-shifted supports the hypothesis that the northern part of CAHA J23056+6016 is slightly tilted towards the observer. A red counterpart to this blue-shifted emission is not detected in the SMA data. This supports indirectly the hypothesis that the additional CO emission lines seen at lower velocities in the SMT spectrum in Figure~\ref{smt-spectrum} are caused by foreground material and are not directly related to CAHA J23056+6016 because the emission is filtered out in the interferometric observations with the SMA. 

C$^{18}$O(2--1) emission is overplotted in blue and red contours in Figure~\ref{mm-image}. Although only detected in two adjacent velocity channels there is a clear spatial offset between the two peaks roughly in east-west direction and thus parallel to the dusty disk and perpendicular to the outflow mentioned above. These data, thus, seem to probe the gaseous counterpart to the dusty disk and the shift of the peak in velocity space can be interpreted as sign of rotation within the gas. It is interesting to note, that the gaseous disk component appears to have a larger physical extent than the dusty component probed in the continuum observations.    However, there are certainly limitations to the current data set and the interpretation: While the red component of the emission appears elongated as expected for a rotating disk, the blue component is in addition much more extended in north-south direction. Furthermore the peak emission in the red and blue component is not perfectly aligned with the dark lane seen in the NIR images. Still, the C$^{18}$O(2--1) data support the picture of an object that is surrounded by a gaseous, rotating disk even if observations with higher sensitivity and spectral/spatial resolution are certainly eligible. 


\subsubsection{Disk mass estimate from C$^{18}$O emission}\label{gasmass}
The C$^{18}$O(2--1) emission can also be used to estimate the related H$_2$ column density and from this the related gas mass, if the distance to the object is known. \citet{wilson2009} derive the H$_2$ column density via
\begin{equation}
N({\rm H_2})=2.65\cdot10^{21}\int T\,{\rm d}v=2.65\cdot10^{21}\cdot I \;[{\rm cm^{-2}}]
\end{equation}
where $I$ is the integrated emission in K$\cdot$km/s. In the Rayleigh-Jeans approximation we find $I=6.86$ K$\cdot$km/s which translates into
\begin{equation}
N({\rm H_2})\approx1.82\cdot10^{22}\;{\rm cm^{-2}}
\end{equation}
The mass of the gas can then be computed using
\begin{equation}
M_{gas}=N({\rm H_2})\cdot A \cdot m_{\rm H} \cdot \mu\quad [{\rm g}]
\end{equation}
with $A$ being the size of the emitting region, $m_{\rm H}$ being the atomic mass unit, and $\mu$ is the mean molecular weight of the gas. To estimate the size of the emitting region we averaged the sizes of the blue and red contours in Figure~\ref{mm-image} and found $\sim$40 arcsec$^2$. Furthermore, we use $m_{\rm H}=1.66\cdot 10^{-24}$ g and $\mu=2.7$ to account for roughly 35\% helium in the gas. With 1 M$_{\sun}\approx 1.99\cdot 10^{33}\;{\rm g}$ we derive
\begin{equation}
M_{gas}\approx 4.5\;{\rm M}_\sun\quad.
\end{equation}
This figure is at least a factor of 2 larger than the values we derived for the disk mass from the dust continuum data in section~\ref{disk_mass_section}. However, as mentioned above and seen in Figure~\ref{mm-image}, the gas emission is also more extended than the compact and unresolved dust continuum emission. Together these estimates show that up to a few solar masses of gas and dust are confined in the immediate circumstellar environment of CAHA J23056+6016.

Although frequently done in the literature, we refrain from estimating the enclosed dynamical mass of the system by equating the gravitational force with the centrifugal force with the latter one being based on the observed rotation velocity and radius of the gaseous disk. This method implicitly assumes a Keplerian rotation of the disk which is no longer valid if the disk mass is not negligible compared to the central mass, which could be the case for CAHA J23056+6016. In addition, there are examples for massive, rotating disk-like structures that explicitly show non-Keplerian rotation \citep[e.g.,][]{beutherwalsh2008}.

\subsubsection{HCO$^+$/H$^{13}$CO$^+$ excitation temperature, column densities and abundances}\label{hco+}
With the IRAM 30\,m telescope we searched for HCO$^+$/H$^{13}$CO$^+$ signatures which are known to trace regions of higher density. The critical density for the (1--0) transition at a temperature of 10 K is roughly $2\cdot10^5$ cm$^{-3}$ \citep{evans1999}. Figure~\ref{hco+-data} shows that both lines, HCO$^+$(1--0) and H$^{13}$CO$^+$(1--0), were detected. At the 
HCO$^+$(1--0) frequency, we find two lines separated by ca.~2.5 km/s. The stronger component
is centered roughly at the velocity found for C$^{18}$O in our SMA data. 
The much weaker H$^{13}$CO$^+$(1--0) line coincides also with the stronger HCO$^+$(1--0) line at around -51.5 km/s. The nature of
the weak HCO$^+$ emission at -49 km/s is unclear. Its relation to CAHA J23056+6016 is unknown, and it might be a foreground
contribution from material not directly associated with CAHA J23056+6016. Since we have not mapped the emission, we 
cannot further investigate this hypothesis at this point, but also in the CO(3--2) spectrum there are additional emission components (see, Figure~\ref{smt-spectrum}). The -51.5 km/s emission on the other hand is most likely related to CAHA J23056+6016 and demonstrates that dense molecular gas is associated with the object. The line strength ratio 
of the two isotopologues ($\sim$ 6:1) indicates that the HCO$^+$(1--0) emission is optically thick as typically the abundance ratio between $^{12}$C and $^{13}$C is in the order of 60:1. Noteworthy are the narrow linewidths of 0.7\,--\,1.0 km/s. Hence, the level of turbulence (potentially 
causing much broader lines) seems to be moderate at best.

Following \citet{purcell2006} we calculate an excitation temperature $T_{ex}$ for both lines and afterwards the column density for the optically thin H$^{13}$CO$^+$ emission. Assuming that the HCO$^+$ line is optically thick ($1-e^{-\tau}\approx1$), as suggested from the line ratios, we can calculate the corresponding $T_{ex}$ according to 
\begin{equation}
T_{ex}=\frac{h\nu_{{\rm u}}}{k}\Bigg[{\rm ln} \Bigg( 1+\frac{(h\nu_{{\rm u}}/k)}{T_{{\rm r}}+J_{\nu}(T_{{\rm bg}})}\Bigg)\Bigg]^{-1}\;[{\rm K}]\quad .
\end{equation}
Here, $h$ is the Planck constant, $k$ is the Boltzmann constant, and $\nu_{{\rm u}}$ is the rest frequency of the transition (see, Table~\ref{lines}). Furthermore, $T_{{\rm r}}$ is the beam corrected brightness temperature for HCO$^+$
\begin{equation}
T_{{\rm r}}=\frac{T_{{\rm antenna}}}{\eta_{{\rm b}}\cdot f}\;[{\rm K}]
\end{equation}
with $T_{{\rm antenna}}$ being the antenna temperature (0.5 K), $\eta_{{\rm b}}$ being the beam efficiency (0.85) and $f$ being the assumed beam filling factor (0.05)\footnote{Based on the extent of the C$^{18}$O emission in Figure~\ref{mm-image} we assume that the HCO$^+$ emission comes from within 6.5$''$. With a beam size for the HCO$^+$ observations of roughly 29$''$ we obtain a filling factor of 0.05.}. Finally, we have
\begin{equation}
J_{\nu}(T_{{\rm bg}})=\frac{h\nu_{{\rm u}}}{k}\frac{1}{e^{h\nu_{{\rm u}}/kT_{{\rm bg}}}-1}
\end{equation}
with $T_{{\rm bg}}$ as the temperature of the background radiation (2.73 K). Plugging in the corresponding numbers we find for the excitation temperature 
\begin{equation}
T_{ex}\approx13.8\;{\rm K}
\end{equation} 
and probably we can assume that the HCO$^+$ and H$^{13}$CO$^+$ emission arises from the same gas and have hence the same excitation temperature. 

Knowing the excitation temperature, we can estimate the optical depth of the optically thin line emission of H$^{13}$CO$^+$ and also the column density of the molecule. 
The optical depth can be derived via
\begin{equation}
\tau_{thin}=-{\rm ln}\Bigg[1-\frac{T_{{\rm r}}}{T_{ex}-J_{\nu}(T_{{\rm bg}})}\Bigg]
\end{equation}
where $T_{{\rm r}}$ is now the beam corrected brightness temperature for H$^{13}$CO$^+$ for which $T_{antenna}=0.08$ K. Beam efficiency and filling factor are the same as above. We find
\begin{equation}
\tau_{thin}\approx0.15\quad.
\end{equation}

Assuming LTE conditions the total column density $N$ can be estimated using
\begin{equation}
N =N_{\rm u}\frac{Q(T_{ex})}{g_{\rm u}}\frac{1}{e^{-E_{\rm u}/kT_{ex}}}\;[{\rm cm^{-2}}]
\end{equation}
where $N_{\rm u}$ is the column density of the upper state, $g_{\rm u}$ the degeneracy of the upper state, $E_{\rm u}$ the energy of the upper state, and $Q(T_{ex})$ the partition function. The upper-state column density can be computed via
\begin{equation}
N_{\rm u}=\frac{8\pi k\nu^2_{\rm u}}{hc^3A_{\rm ul}}\int^{\infty}_{-\infty}T_{\rm b}{\rm d}\nu\;[{\rm cm^{-2}}]
\end{equation}
with, again, $k$ and $h$ being the Boltzmann and the Planck constant, respectively, $A_{\rm ul}$ being the Einstein coefficient of the transition in s$^{-1}$, $c$ being the speed of light, and $\nu_{\rm u}$ and $\int^{\infty}_{-\infty}T_{\rm b}{\rm d}\nu$ being the transition frequency and the measured line intensity, respectively (see, Table~\ref{lines}). The partition function $Q(T_{ex})$ for linear rotators can be approximated by
\begin{equation}
Q(T_{ex})\approx\frac{kT_{ex}}{hB}
\end{equation}
\citep{purcell2006} where $B$ is the rotational constant, i.e., 43377 MHz for H$^{13}$CO$^+$ \citep{schmid-burgk2004}. Plugging in the derived value for $T_{ex}$ and using $g_{\rm u}=2J+1=3$, $A_{\rm ul}=3.8534\cdot{\rm 10}^{-5}$s$^{-1}$, and $E_{\rm u}=4.16$ K we derive
\begin{equation}
N({\rm H^{13}CO^+})\approx 7.0\cdot 10^{10}\;{\rm cm^{-2}}
\end{equation}
for the H$^{13}$CO$^+$ column density yielding $N({\rm HCO^+})\approx 4.2\cdot 10^{11}\;{\rm cm^{-2}}$ assuming an abundance ratio of 60:1 for $^{12}$C/$^{13}$C.

To get an estimate of the overall HCO$^+$ and H$^{13}$CO$^+$ abundance we need to derive the H$_2$ column density. In section~\ref{gasmass} we already estimated the H$_2$ column density based on the interferometer data. For a direct comparison with the HCO$^+$/H$^{13}$CO$^+$ IRAM 30\,m observations it is, however, more appropriate to derive the column density also from single dish data. This can be done using the dust continuum emission we observed with the IRAM 30\,m telescope as here the beams of the dust continuum and the HCO$^+$/H$^{13}$CO$^+$ observations are the most comparable. It holds
\begin{equation}\label{NH2}
N({\rm H_2})=\frac{F\cdot R_{g/d}}{\Omega\cdot B_{\nu}(T_{dust})\cdot\kappa_{\nu}\cdot m_{\rm H}\cdot\mu}\;[{\rm cm^{-2}}]
\end{equation}
\citep[see, e.g.,][]{vasyunina2009}, where $F$ is the dust continuum peak emission, $R_{g/d}$ is the assumed gas-to-dust ratio, $\Omega$ is the solid angle (in steradians) covered by the beam, $B_{\nu}(T_{dust})$ is the Planck formula for the assumed black body emission of the dust at frequency $\nu$ and temperature $T_{dust}$, $\kappa_{\nu}$ is the dust opacity, $m_{\rm H}$ is the atomic mass unit, and $\mu$ is the mean molecular weight of the gas. For $F$ we use $12.9\cdot10^{-26}$ erg/s/cm$^2$/Hz (see section~\ref{disk_mass_section}). Furthermore, we use $R_{g/d}$=100 and $\kappa_{\nu}$=1.0 cm$^2$g$^{-1}$. $m_{\rm H}$ and $\mu$ are the same as in section~\ref{gasmass}. The beam solid angle can be computed via
\begin{equation}
\Omega=2\pi\Big(1-\cos \frac{\omega}{2}\Big)[{\rm sr}]
\end{equation}
with $\omega$ being the beam size in radian\footnote{1 rad $\approx$ 206265 arcsec}. Although the beam size for the IRAM30 dust continuum observations was $\sim$10.5$''$, we have again to take into account the assumed size of the H$^{13}$CO$^+$ emission. Thus, to be consistent, we use an effective beam size of 6.5$''$. Based on equation~\ref{NH2} we hence derive a hydrogen column density of
\begin{equation}
N({\rm H_2})\approx1.32\cdot10^{22}\;{\rm cm^{-2}}
\end{equation}
yielding an abundance ratio for H$^{13}$CO$^+$ of
\begin{equation}
X=\frac{[{\rm H^{13}CO^+}]}{[{\rm H_2}]}=\frac{N({\rm H^{13}CO^+})}{N({\rm H_2})}\approx5.3\cdot10^{-12}\quad.
\end{equation}
We will discuss these results below in section~\ref{discussion_chemical_abundances} but note already that the H$_2$ column density derived here is in general agreement with the value based on the C$^{18}$O SMA data (section~\ref{gasmass}).

\subsubsection{Detection of C$_2$H}\label{c2h_section}
In Figure~\ref{ch-data} we show the C$_2$H(2--1) emission line towards CAHA J23056+6016 detected with IRAM30. The line is weak and very narrow (Table~\ref{lines}) but clearly detected and the observed velocity agrees with the assumed velocity of rest of CAHA J23056+6016 of -51.3 km/s. The possible origin and the interpretation of the emission will be further discussed in section~\ref{discussion_chemical_abundances}.

\section{Radiative transfer modeling}
The central illumination source in CAHA J23056+6016 is obscured in
the NIR and MIR by the
circumstellar material, so that radiative transfer
calculations are necessary to analyze the radiation that we
receive from the dust particles near the source.

Due to the scattering integral in the radiative transfer equation and
the complex geometry of the dust configuration around the source, 
calculations of NIR scattered light images are time-consuming. 
Given the many free parameters that
are necessary to describe the disk distribution, the radiation source,
and the outflow, an automated fitting procedure and an extensive
exploration of the parameter space are therefore 
prohibitive.

Moreover, the IRAC images, especially at 5.8 $\mu$m, are 
expected to be influenced by PAH emission which opens up an even
wider parameter space.

For this paper, reporting on the detection of the new large, massive disk, we restrict the analysis to a simple model based on a 
2D density distribution for the disk and the reflection nebula.

From the images, it is obvious that a 2D modeling will not able to
reproduce the warping of the disk and the $H$ and $K$-band 
flux maximum in the left part
of the nebula over the dark lane. The model was chosen to estimate
the disk and envelope parameters that are needed to reproduce the dark lane
and the overall flux pattern in the reflection nebula.

The models make use of the constraints derived from the SMA image
analysis, namely a total disk mass between 0.4 and 2.7 M$_\odot$
(see Table 2), and a radial disk extent $\le$\,7000 AU (see section 3.3.2.).
For the density distribution of the dust in the disk, we use a
parameterized disk model with a radial power-law profile and an 
exponential disk atmosphere \citep[e.g.,][]{pascucci2004}
of the form
\begin{equation}
n(r,z) =
n_0
\left( \frac{r}{r_0} \right)^\alpha
\exp\left[ -\left( \frac{z}{rh}\right)^2\right]
\end{equation}
This dust number density is normalized by $n_0$, assumes a clear inner zone around
the central object up to the inner radius $r_0$,
a truncation at the outer radius $r_1$, the radial powerlaw exponent
$\alpha$, and the $e$-folding scale height ratio $h=H/R$
($H$ is the $z$-value for which the density has dropped by 1/$e$ for a given radius).
The disk may be inclined with respect to the edge-on view by the angle $i$,
and rotated by $\delta$ in the plane-of-sky relative to the north. 

To obtain a reasonable description of the boundaries of the reflection
nebula, we normalized the $H$ and $K$ image equally and added them, rotated
the resulting image
to achieve a vertical symmetry axis, and added the image parts left and
right from the axis.
In this image, the border line of the nebula was fitted with a 4th-order 
polynomial.
Figure~\ref{envelope_edge} shows the resulting image as inlet and the contour plot with the
fitting polynomial.

For the density structure in the envelope, we have used several model
types including a thin sheet at the nebula border and a density 
distribution filling the entire cavity. The finally used distribution
has a maximum at the nebula border and a linear decrease toward the symmetry
axis down to a constant density, as well as a linear decrease along this axis.
The distributions were attached at a height over the disk where the 
nebula border line height is equal to the disk scale height for the outer radius.
The dust particles were assumed to be 0.1 $\mu$m-sized spherical silicate
grains using the optical properties described in \citet{btdraine2003}.

We have varied the luminosity of the central source to cover the range
from young, intermediate mass objects \citep[Herbig Ae/Be stars,][]{herbig1960} to massive stars.
The radiative transfer calculations were performed using a ray-tracing
code \citep{steinacker2006}.

\subsection{Disk properties}
Figure~\ref{disk_model} shows the observed $K$-band image (left) and the synthetic image 
based on a disk density model (right) with a dust number density
of $10^6$ m$^{-3}$ at the inner disk radius $r_0=$ 20 AU, a density
powerlaw index of 1.8, a disk inclination of 10$^\circ$, a nebula tilting 
of 10$^\circ$ in the plane-of-sky, an outer radius of 4000 AU, and a disk
scale height ration of 0.2. The outer radius in the model indicates that indeed the true disk size is smaller than inferred from the NIR images and that shadowing effects may play a role. The model was chosen by comparing the resulting image with the model
images of about 400 other models varying the disk parameters. 
We have restricted the modeling to the $K$-band image which shows the
best signal-to-noise ratio.

It reproduces the position, tilting, extent, and thickness of the dark lane.
The asymmetric darkening in the right part of the disk is not reproduced
within the 2D model. The images are not sensitive to changes in the
inner or outer disk radius since the disk is seen almost edge-on and
the outer disk parts are shadowed by the inner parts.
The images are sensitive to the disk inclination since a larger inclination
amplifies the scattered light flux from the upper disk part in disagreement
with the observed flux pattern. Moreover, for inclinations $>$\,15$^\circ$, 
the disk lane is stronger bend than observed.
The model images also failed to reproduce the $K$-band image when the disk
scale height deviated by more than 0.1 from the value 0.2.
There was no strong dependence on the radial spectral index.

Since the outer disk parts do not modify the images strongly as long as
the disk is kept almost edge-on, a variation in the disk mass with
otherwise unchanged disk properties was possible within the mass range 
determined from the SMA data without changing the images in their
overall appearance.

The image shows stronger scattered light flux from the upper disk
atmosphere compared to the flux from below the dark lane due to the
inclination of the disk.

A clear difference between the observed and the synthetic image
is the lower part of the reflection nebula which is much weaker than
the upper part in the observed flux while the theoretical image shows
comparable fluxes in the outer nebula parts.
We have investigated if forward scattering or disk inclination could cause
this effect. The scattering phase function is too homogenous to
reduce the flux in such a way, and a stronger disk inclination will
produce too much flux from the upper reflection nebula.
Since this part 
shows also a stronger deviation from rotational symmetry in the observed
image, we speculate that outer material may cause extinction patches
which reduce the flux of the lower nebula. 

\subsection{The central source}
A determination of the parameters of the central source is hindered
by the ambiguity in the explanation of the scattered light flux from
the nebula. The same flux value can be obtained when the stellar flux
is decreased but the column density within the nebula is increased.
Moreover, the dust opacities are poorly known, and will add another
factor of 5 of uncertainty to the stellar properties determined from
the scattered light measurements.

\section{Discussion}

\subsection{Disks around young stars - masses and sizes}
Putting the results from the previous sections in context with other young objects, the circumstellar disk surrounding CAHA J23056+6016 appears quite exceptional in two ways: (1) it is fairly large with an outer radius $\le$ 7000 AU, and (2) it is fairly massive with a mass in the order of up to a few M$_\sun$ (see, sections~\ref{disk_mass_section} and ~\ref{gasmass}). In particular for the second point we emphasize that our SMA observations probe almost exclusively emission from the disk and at least the mm-continuum data are not contaminated by additional flux from a remnant envelope. 

Circumstellar disks around young, low-mass stars (TTauri stars) typically have masses in the order of a few times 10$^{-3}$--$10^{-2}$ M$_\sun$ and they have a maximum radial extent of a few hundred AU \citep[e.g.,][]{beckwith1990,koerner1995,andrewswilliams2005}. In these systems the disk-to-star mass ratio is roughly 0.5--1\% although the ratio is certainly larger at earlier evolutionary stages when mass is continuously fed to the disk by a circumstellar envelope. For instance, \citet{eisner2006} found in a mm-survey in the Trapezium cluster the mean circumstellar disk mass to be in the order of the minimum-mass solar nebula ($\sim$0.005\,--\,0.01 M$_\sun$), but they found a few sources with disk masses between 0.13\,--\,0.39 M$_\sun$ (with at least a factor of 3 uncertainty) for which it seems likely that they are in an earlier evolutionary stage.

For young Herbig Ae/Be stars the derived disk masses range from values also comparable to the minimum-mass solar nebula \citep[e.g.,][]{mannings1997,thi2004,alonso-albi2009} to slightly higher masses, i.e. 
$\sim 10^{-1}$ M$_\sun$ \citep[e.g.,][]{henning1998}. The sizes of these disks appear to be in the same range as those around lower mass stars or only slightly larger 
\citep[e.g.][]{mannings1997,dutrey2007,schreyer2008,alonso-albi2009}. There are at least two noteworthy exceptions, though: The first one is the disk around the pre-main sequence A-type star IRAS 18059-3211 where the gas disk extends out to 1800\,--\,3000 AU and where the disk mass is estimated to be in the order of 0.3\,--\,0.9 M$_\sun$ as derived from dust continuum data \citep{bujarrabal2008,bujarrabal2009}. Unfortunately the distance to this source is not extremely well constrained leaving these numbers with comparatively high uncertainties. In addition, an applied disk model predicts a disk mass that is an order of magnitude lower than the values cited above. The second one, is the heavily discussed circumstellar disk in M17 \citep[e.g.,][]{chini2004,sako2005,steinacker2006,nurnberger2007,nielbock2008}. Most recent results suggest that the central object is an intermediate mass star between 3\,--\,8 M$_\sun$ \citep{nielbock2008}, but the disk was not detected with the SMA and its mass and actual extent depend strongly on modeling assumption \citep{steinacker2006}.

For high-mass stars ($\ge$ 8 M$_\sun$), as mentioned already in the introduction, the existence of circumstellar disks has not been fully established although there is growing observational evidence that they exist \citep[e.g.,][]{cesaroni2007,reid2008,jiang2008,beuther2009}. For example, \citet{schreyer2006} found evidence for a rotating $\sim$1.5 M$_\sun$ disk structure around the 8-10 M$_\sun$ object AFGL 490 based on interferometric observations of C$^{17}$O line emission. 
However, more often than not the main problem is the lack of spatial resolution which would help to disentangle rotating "disk-like" structures, "pseudodisks" \citep{gallishu1993}and gaseous tori (possible remnants from the parental molecular cloud core) from real accretion disks surrounding and feeding material to the young massive star. But even with sufficient spatial resolution, e.g., using MIR interferometry around 10 $\mu$m where dusty disks are bright, so far the existence of large and/or massive circumstellar disks around massive young stellar objects could not be proven \citep{linz2009,dewit2009}. Only recently \citet{okamoto2009} found direct evidence for a flared disk around the 10 M$_\sun$, young star HD200775 from direct MIR images. This radius of the disk extends out to 1000 AU, but the mass, derived from unresolved emission at 350 GHz, is only a few 10$^{-2}$ M$_\sun$, with the exact value depending on the assumed dust opacities and temperature. Indirect evidence for accretion disks around young objects of similar mass has been found by \citet{boley2009} and \citet{davies2009} for S 235 B and W33A, respectively. Resolved images of the latter two disks have not yet been obtained.

This shows, that CAHA J23056+6016 is a good candidate for being among the largest and most massive circumstellar disks detected and resolved on multi wavelengths observations. However, one should keep in mind that it has not yet been shown that the disk is indeed an accretion disk from which matter is transferred onto the central object while angular momentum is transported outwards. The initial concept of "pseudodisks" actually predicts gaseous, flattened structures with sizes at least comparable to what is observed here \citep{gallishu1993}. Interestingly, however, with the current instrument sensitivities the dusty disk component of CAHA J23056+6016 appears to be more compact than the detected gaseous disk component (section 3.3.3, Figure~\ref{mm-image}). It would be interesting to check observationally for signs of accretion coming from a compact inner accretion disk that is embedded in a somewhat larger flattened molecular structure.

\subsection{Mass of the central object}
Our current data does not allow us to constrain the mass of the central source empirically. Only indirectly, by comparing the derived disk mass to those of other systems as done above, we can speculate that most likely  the central source of CAHA J23056+6016 is at least an intermediate mass object with several solar masses. Trying to derive the spectral type of the central source from NIR spectroscopy of the scattered light in the reflection nebula could possibly be a means of getting more direct information of the object.  


\subsection{CAHA J23056+6016 - an evolved object?}
Up to now it was assumed that CAHA J23056+6016 is a \emph{young} stellar object. But what makes us believe that it is indeed young and not evolved? At first glance, the NIR morphology of CAHA J23056+6016 is not only compatible with a young object but does also resemble  that of proto-planetary nebula (PPNe), i.e., post-AGB stars \citep[e.g., see][for high-resolution, NIR HST images of OH 231.8+42]{meakin2003}. In addition, also for the intermediate mass, pre-main sequence object IRAS 18059-3211 (already mentioned above) \citet{ruiz1987} initially suggested that the object is rather an evolved source and not a pre-main sequence star.

We note that despite the apparent similarities in the morphology compared with evolved stars, CAHA J23056+6016 is lacking some crucial properties of older objects. (1) \citet{meakin2003} found a very filamentary sub-structure in the bipolar cones surrounding the PPNe OH 231.8+42. This structure is caused by material that is ejected from the central source. This mass loss is, however, neither constant nor homogenous in space or time and leading to knots and filaments in the bipolar cones. Such a sub-structure is not seen in the nebula around CAHA J23056+6016 (Figure~\ref{nir-images}). The nebula around CAHA J23056+6016 appears rather smooth and its brightness distribution can be very well explained with an illuminated outflow cavity with small dust grains scattering the light coming from the central source. (2) The mass loss of evolved stars is typically accompanied by high-velocity molecular outflows and jets \citep[e.g.,][]{sanchez2004} and the observed outflow velocities of PPNe typically exceed several tens of km/s with respect to the system's velocity of rest \citep{bujarrabal2001}. But, again, we do not find any evidence for a collimated jet or a high-velocity outflow component for CAHA J23056+6016, neither in the NIR images (where shock-induced hydrogen emission could show up in the $K'$-band image) nor in the CO data. (3) Finally, evolved objects also show expanding molecular tori lying in the disk plane \citep[e.g.,][]{hirano2004,meakin2003}. In the case of CAHA J23056+6016, however, the molecular gas that we see in our C$^{18}$O data is well confined to a small velocity regime making the interpretation of a rotating, disk-like structure much more likely compared to an expanding torus. 

Based on these findings and the fact that we are confident that CAHA J23056+6016 is indeed a young and not an evolved object. 
 
\subsection{An isolated massive circumstellar disk} 
In one way CAHA J23056+6016 appears to be different compared to the majority of young intermediate mass and higher mass objects. Most of the young, sources mentioned above happen to be in young clusters or star-forming regions (e.g., M17, HD200775, S 235 B, W33A, as well as the lists of massive protostellar disk candidates in \citet{reid2008} and \citet{cesaroni2007}). For massive stars, \citet{smith2009} (among others) claim that they can only form in cluster environments and that it is the overall evolution of the cluster which leads to the formation of massive objects in the center eventually. Examples for this can be found in Figure~\ref{color-image-wide}, where one can nicely see how the young, more massive stars in the HII regions are surrounded by a noticeable group of additional objects. The sizes of these clusters are in very good ag-reement with the findings from \citet{ladalada2003} who found the typical size of young stellar cluster to be in the order of 0.5\,--\,1 pc which corresponds to roughly 0.5\,--\,1$'$ in Figure~\ref{color-image-wide} for a distance of 3.5 kpc. In this context it is even more striking that CAHA J23056+6016 is rather isolated while in other regions of the same molecular cloud complex star formation occurs in a cluster-mode.

One interesting exception to the rule appears to be the intermediate mass, pre-main sequence star IRAS 18059-3211, which, as noted above, was initially classified as an evolved and not a young source. Indeed, the vicinity of CAHA J23056+6016 looks very similar to that of IRAS 18059-3211 as there is no direct evidence for a surrounding young cluster (Figure~\ref{color-image-wide} and Figure~\ref{color-image-zoom}). Based on comparison to models for stellar evolution of low-mass stars \citep{baraffe1998}, we estimate that in our deep NIR images we are able to detect objects with masses below 0.1 M$_\sun$ for an assumed age of 1--5 Myr at a distance of 3.5 kpc even if we assume a line-of-sight extinction of $A_V$=15 mag. However, we do not find direct evidence for a (tight) cluster of young sources of lower mass. Also, the reddish nebula south of the brightest star lying west of CAHA J23056+6016 in Figure~\ref{color-image-zoom} does not harbor any young embedded source:  We did not detect this object in direct mm-continuum observations with the IRAM 30 m telescope carried out in the same night as the observations of CAHA J23056+6016 (see section~\ref{iram_obs}). We rather assume that the nebula is reflective in nature and associated with the brighter star lying north of it. This star and the bright object east of CAHA J23056+6016 in Figure~\ref{color-image-zoom} are the brightest NIR source in the direct vicinity of CAHA J23056+6016. Based on their NIR colors \citep{cutri2003} we estimate their spectral types to be K2-K4 \citep{ducati2001} which would make them foreground objects compared to CAHA J23056+6016. Thus, even if we cannot exclude that maybe multiple low-mass sources are hidden behind the dusty disk and/or the nebula around CAHA J23056+6016, we currently have no indications for a young cluster within 0.5\,--\,1 pc (the typical cluster size)  of CAHA J23056+6016. If this was the case, a solid determination of the central mass would be crucial. As mentioned above, in one theory more massive objects are thought to form mostly (if not solely) in company with other objects \citep{bonnell2006,smith2009}. The detection of CAHA J23056+6016, if it was a massive object, would talk very much in favor for a star formation process for massive stars that is very similar to that for low-mass object \citep{mckee2003, krumholz2009}: Single objects can form from the collapse of a molecular cloud core, and to reach its final mass an object accretes material from a surrounding circumstellar disk. 

\subsection{Chemical abundances}\label{discussion_chemical_abundances}
\subsubsection{H$^{13}$CO$^+$ and HCO$^+$}
In section~\ref{hco+} we derived an H$^{13}$CO$^+$ abundance of roughly $5.8\cdot 10^{-12}$ which yields an HCO$^+$ abundance of $3.5\cdot 10^{-10}$ assuming a $^{13}$CO/$^{12}$CO ratio of 60:1. These values can be compared to those found in young cores and envelopes and also circumstellar disks. 

For two massive embedded protostars in the IRAS 18151-1208 region \citet{marseille2008} found an H$^{13}$CO$^+$ abundance of 3~--~$8\cdot10^{-11}$. Similarly, \citet{zinchenko2009} found for their census of regions of massive star formation values between $5\cdot 10^{-11}$ and $3\cdot10^{-10}$. Interestingly, for the envelopes of low-mass protostars \citet{vankempen2009} found values that are roughly two to three orders of magnitudes higher (2~--~$8\cdot10^{-8}$). \citet{thi2004}, however, found for the solar-mass protostar IRAS 16293--2422 an H$^{13}$CO$^+$ abundance of $\sim 2\cdot 10^{-11}$ which is similar to the values mentioned above for high-mass protostars. The same authors provide also H$^{13}$CO$^+$ abundances for circumstellar disks around low-mass stars. Here, the values appear to be even lower, although a large dispersion exists. For 3 low-mass disks, values between $\sim 3\cdot 10^{-13}$ and $\sim 3\cdot 10^{-11}$ were found. In addition, for 2 disks around Herbig Ae/Be stars \citet{thi2004} give HCO$^+$ abundances of $1\cdot10^{-10}$ and $\sim 8\cdot10^{-12}$. 

Given the large spread in these values, it seems difficult to derive any firm conclusion about the scientific meaning of the abundances we find for CAHA J23056+6016. In addition, the production and depletion of the molecules depend strongly on the environment \citep{thi2004}. HCO$^+$ is mainly produced by ${\rm H}_3^+ + {\rm CO} \rightarrow {\rm HCO}^+ + {\rm H}_2$ in a gas phase reaction and here the formation is, among other parameters, dependent on the surrounding ionization degree. For the depletion, there are in general different mechanisms possible, e.g., dissociative recombination (${\rm HCO}^+ + e^- \rightarrow {\rm CO} + {\rm H}$), reaction with water (${\rm HCO}^+ + {\rm H}_2{\rm O}\rightarrow {\rm H}_3{\rm O}^+ + {\rm CO}$) or charge transfer with low ionization potential elements such as Na or Mg \citep{nikolic2007}. Again, these reactions depend on the environmental conditions. 

Despite all these uncertainties, we can note at least two things: First, the abundances derived for CAHA J23056+6016 lie in the same range as those found in circumstellar disks around young, low-mass or intermediate mass stars, but seem lower than the values found in (high-mass) protostars. Second, using the derived abundances as lower limit for the ionization fraction in the disk, we find that the value is high enough to cause magnetorotational instabilities in the disk which then drives turbulence and mixing \citep[e.g.,][]{balbus2009}. Hence, we can expect a certain amount of disk dynamics to occur and possibly also accretion if the central source features a magnetic field that suffices to reach and tap the ionized parts of the disk. 

\subsubsection{C$_2$H}
As seen in section~\ref{c2h_section}, we detected C$_2$H emission towards CAHA J23056+6016. 
The detection of this molecule and its interpretation in the astrophysical context of star formation is subject of very recent publications in the literature.

\citet{beuther2008} found that C$_2$H was present in a larger sample of massive star-forming regions covering various evolutionary stages in the star formation process. They suggest that the molecule could potentially be used to trace the initial conditions of massive star formation, as in the earliest phases its abundance remains high due to constant replenishment of elemental carbon from CO being dissociated by the interstellar UV photons. In later stages, the molecule is transformed into other species in the dense protostellar core regions and might only remain traceable in the outer core regions where it is excited by external UV radiation. 

In circumstellar disks, there is only evidence of C$_2$H in three cases (DM Tau, LkCa15, and MWC 480; Henning et al. 2010, submitted). However, the detection of the molecule in the disk around the Herbig Ae star MWC 480 was very weak. From their observations and modeling the chemical evolution of the disks the authors concluded that C$_2$H is indeed a sensitive tracer of the UV radiation field and that the stellar UV luminosity affects the evolution of C$_2$H in the disk globally, while X-ray radiation appears to be more important in the inner disk regions close to the star. In this way, the lack of C$_2$H in the Herbig Ae disk compared to the TTauri disks is explained with the higher UV flux from the central source. 

 
In this context, we suggest that the C$_2$H we detect towards CAHA J23056+6016 might be another hint for the young age of the source. However, the current observations lack the required spatial resolution to conclude where the emission is coming from. Apart from emission from the disk itself  it seems possible that the  C$_2$H flux is coming from the outer and thus externally illuminated parts of the disk / nebula system and not necessarily from the inner regions. In this case the nearby HII regions (Figure~\ref{color-image-wide}) could be the source of the required UV photons to excite the emission. 


\section{Summary and conclusions}
In this paper we presented a new large and massive circumstellar disk (named CAHA J23056+6016) presumably surrounding a young stellar object. Our findings can be summarized as follows:
\begin{itemize}
\item The disk is seen almost edge-on and was detected in high resolution NIR images where it caused a dark lane intersecting a bipolar nebula that is most likely illuminated by a young stellar source. The dark lane stretches over $\sim 3''$ which, for an assumed distance of 3.5 kpc, translates into $\sim$10500 AU, but due to shadowing effects the true physical disk size is probably smaller (see also last point). 
\item  CAHA J23056+6016 appears to be one of the largest circumstellar disks known today but at the moment we have no empirical evidence whether it is an actual accretion disk or a "pseudodisk" resulting from the collapse of a moderatly magnetized rotating molecular cloud core or a flattened remnant envelope.
\item The young stellar source residing behind the disk is not detected in the NIR images as the optical depth through the disk is too high. 
\item In \emph{Spitzer IRAC} images the object is also detected and at 5.8 $\mu$m one starts probing directly through the disk and detects flux mainly coming from the central disk regions. 
\item This finding is supported by the fact that the peak of the emission seen in different \emph{Spitzer IRAC} filters is shifting with wavelength and that at the longest available \emph{IRAC} wavelength (5.8 $\mu$m) the emission peak coincides with the center of the dark lane seen in the NIR images.
\item In interferometric mm dust continuum observations with the SMA we detected but not resolved CAHA J23056+6016 at the 11.1 mJy/beam level. This sets the upper limit for the disk radius to $\le$\,7000 AU.
\item The mm dust emission peaks again nicely on the dark lane seen in the NIR images and coincides perfectly with the  \emph{IRAC} 5.8 $\mu$m emission. From the unresolved continuum data we estimate the disk mass to be in the order of 1 M$_\sun$ making it one of the rare cases where a massive disk is resolved in NIR images. 
\item We found a molecular outflow in CO(2--1) coinciding perfectly with the northern part of the bipolar nebula seen in the NIR images. In addition, we found indications for a gaseous disk component in C$^{18}$O data, with the emission being perpendicular to the outflow and showing signs for rotation. The gaseous disk is even larger than that derived from the NIR images or the dust continuum observations and yields a mass of $\sim$4.5 M$_\sun$.
\item We also detected H$^{13}$CO$^+$ and HCO$^+$ line emission from CAHA J23056+6016 which arises from regions of higher density. The derived abundances are comparable to those found in other circumstellar disks but lower compared to protostellar source.
\item The detection of C$_2$H supports a young age for the object. We suggest that the emission could either arise from the disk itself but could also be emitted from externally illuminated regions of the disk / envelope system that are subject to high UV radiation fields from nearby HII regions.
\item The current data does not allow us to constrain the mass of the central object directly but, based of the derived disk mass, we speculate that it could have up to several solar masses as well. 
\item Interestingly, we find no signs of a young cluster around CAHA J23056+6016 within the typical cluster size of 0.5\,--\,1 pc. If the central source were rather massive as well then we had found a rare case of isolated massive star formation.
\item Using a simple 2D dust density profile for the disk and the 
reflection nebula, the radiative transfer modeling of the $K$-band
image could constrain the inclination of the disk to be 5\,--\,15$^\circ$
from an edge-on view. The scale height ratio of the disk ranges
from 0.1 to 0.3 to be consistent with the extent of the dark lane.
The radial power-law index was not as well constrained as the inner
and outer radius of the disk, where the latter parameter is found to be $\sim$4000 AU. 
\end{itemize}
The disk of CAHA J23056+6016 is extremely large and very massive and the fact that we have spatially resolved information at a variety of wavelengths makes this object an ideal laboratory to study the properties of a disk surrounding a young presumably intermediate mass or even massive star. Our finding argues for a star formation process that is functioning over a wide mass range, ranging from sub-stellar objects \citep[e.g.,][]{zapatero2007,quanz2010} to, as it appears, at least several solar masses. In addition, as higher mass stars seem to process their circumstellar disks faster than low-mass objects \citep[e.g.,][]{alonso-albi2009}, CAHA J23056+6016 provides the opportunity to study an early stage of disk evolution where the disk is still massive. 

For future observations of CAHA J23056+6016 it might be interesting to aim at an even higher spatial resolution of the dust continuum emission to eventually resolve the disk. Filling in the SED in sub-mm and MIR wavelengths could in addition help to even better characterize the disk geometry and dust properties. In parallel, observations of the gaseous disk component with higher spectral resolution will shed additional light on the detected rotation signatures. Finally, it might be worth to obtain a NIR spectrum of the reflection nebula. This might allow us to search for spectral features that constrain the spectral type of the central source and possible bear signs of accretion processes.

\acknowledgments
S. P. Q. kindly acknowledges initial support from the German \emph{Friedrich-Ebert-Stiftung}. We are extremely grateful to the staff at Calar Alto observatory, SMT, Subaru telescope and the SMA for the great support during the observations. In particular, we thank Thorsten Ratzka, Carlos Alvarez, Miwa Goto and Usuda Terada for their help during the preparation and execution of the observations. J. S. thanks Christian Fendt, Bhargav Vaidya, and Laurent Pagani for 
helpful discussion.
This research has made use of the SIMBAD database, operated at CDS, Strasbourg, France.  


{\it Facilities:}  \facility{Calar Alto}, \facility{SMA}, \facility{Spitzer}, \facility{IRAM 30}

\clearpage

\begin{deluxetable}{lccllccl}
\rotate
\tablecaption{Results from gaussian fits to molecular lines detected towards CAHA J23056+6016.
\label{lines}}           
\tablehead{
\colhead{Line transition} & \colhead{Rest frequency} & \colhead{Line No.} & \colhead{Line position} & \colhead{Line width}& \colhead{T$_{{\rm mb}}$\tablenotemark{a}} & \colhead{Line intensity} & \colhead{Telescope}\\
\colhead{} & \colhead{[GHz]} & \colhead{} & \colhead{[km/s]} & \colhead{[km/s]}& \colhead{[K]} & \colhead{[K km/s]} & \colhead{}
}
\startdata
CO (J=3--2) 			& 345.795990 & 1 &  -45.54$\pm$0.12 &  1.95$\pm$0.28 &   2.30  &    &SMT \\
					&				& 2&  -49.28$\pm$0.07 &  2.22$\pm$0.16 &   8.39  &  & \\
					&				&3&	  -51.30$\pm$0.07 &   1.54$\pm$0.16 &   6.24 & &\\
					&				& 4& -52.96$\pm$0.04 &    0.97$\pm$0.09 &  -4.93 &&\\	
HCO$^+$ (J=1--0)			& 89.188523	& 1 & -48.92$\pm$0.05 & 1.00$\pm$0.10 &  0.21 & 0.23$\pm$0.02\tablenotemark{b} & IRAM 30\,m \\
					& 				& 2 & -51.48$\pm$0.02 & 1.00$\pm$0.05 &  0.59 & 0.63$\pm$0.03\tablenotemark{b}   & \\
H$^{13}$CO$^+$ (J=1--0)	& 86.754288 	& 1 & -51.44$\pm$0.05 & 0.72$\pm$0.11 &  0.10 & 0.07$\pm$0.01\tablenotemark{b}  & IRAM 30\,m \\
C$_2$H (J=1--0)		&        87.316925 	& 1 & -51.47$\pm$0.04 & 0.69$\pm$0.15 &  0.30 & 0.22$\pm$0.03\tablenotemark{b} & IRAM 30\,m \\
\enddata
\tablenotetext{a}{Main beam brightness temperature}
\tablenotetext{b}{For a main beam efficiency of 0.85 that already includes the forward efficiency}
\end{deluxetable}

\clearpage

\begin{deluxetable}{cccc}
\tablecaption{Estimated disk masses in units of solar masses for different values for the dust opacity $\kappa$ and different dust temperatures. A gas-to-dust ratio of 100 was assumed. 
\label{disk_mass}}           
\tablehead{
\colhead{$\kappa$ at 1.3 mm} & \colhead{$T_{dust}$=20 K} & \colhead{$T_{dust}$=50 K} & \colhead{Ref. for opacity}
}
\startdata
$1.0$ cm$^2$/g: & 2.7 $M_{\sun}$ & 0.9 $M_{\sun}$ &  \cite{ossenkopf1994} \\
$2.3$ cm$^2$/g: & 1.2 $M_{\sun}$ & 0.4 $M_{\sun}$ & \cite{andrewswilliams2005}\\
\enddata
\end{deluxetable}

\clearpage

\begin{figure}
\centering
\epsscale{2.}
\plotone{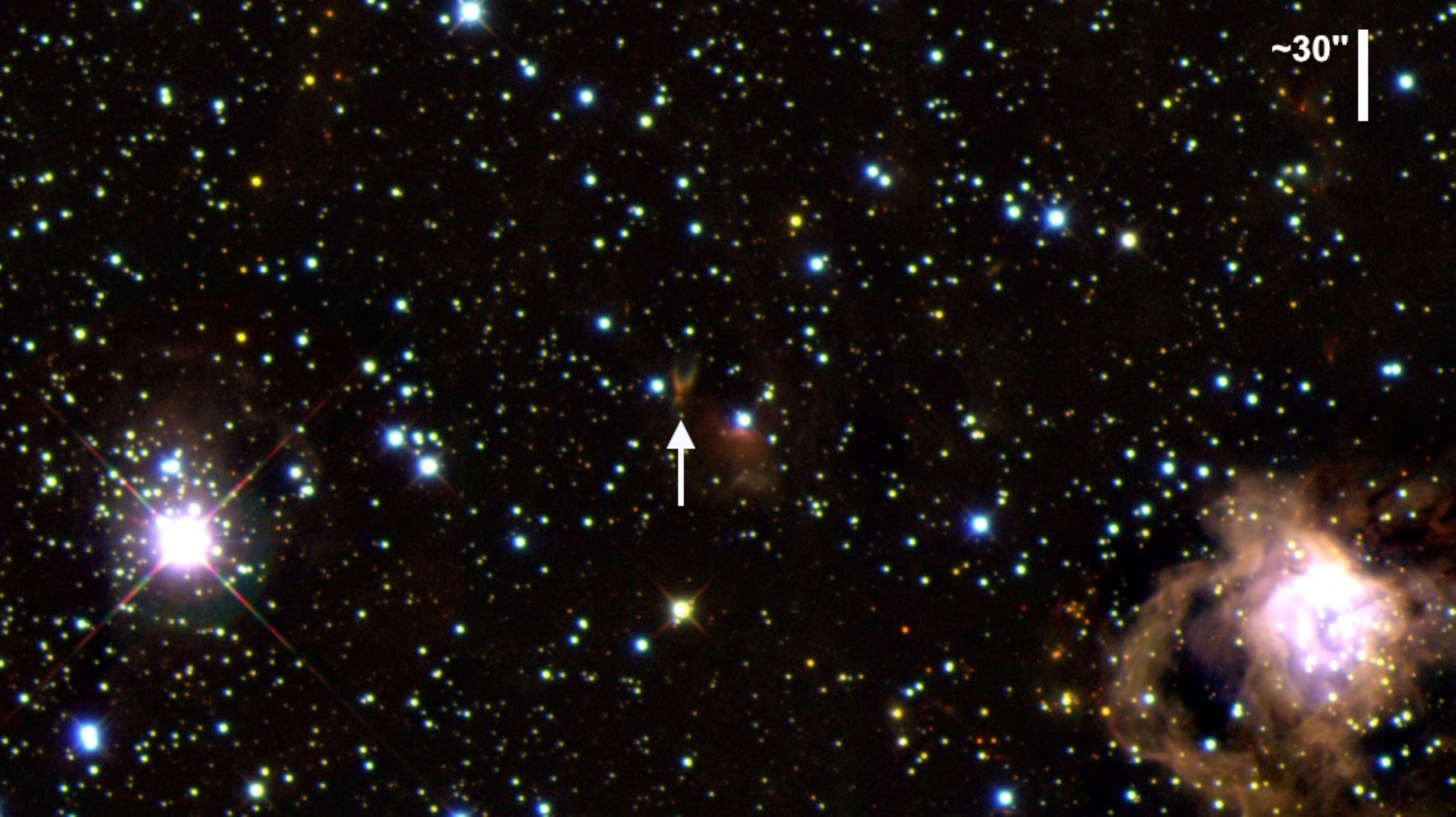}
\caption{NIR color composite consisting of $J$ (blue), $H$ (green) and $K_S$ (red) filter images obtained at Calar Alto with {\sc Omega2000}. The bipolar nebula intersected by a dark lane at the image center is CAHA J23056+6016 (north is up, east to the left). 
\label{color-image-wide}}
\end{figure}

\clearpage

\begin{figure}
\centering
\epsscale{2.} 
\plotone{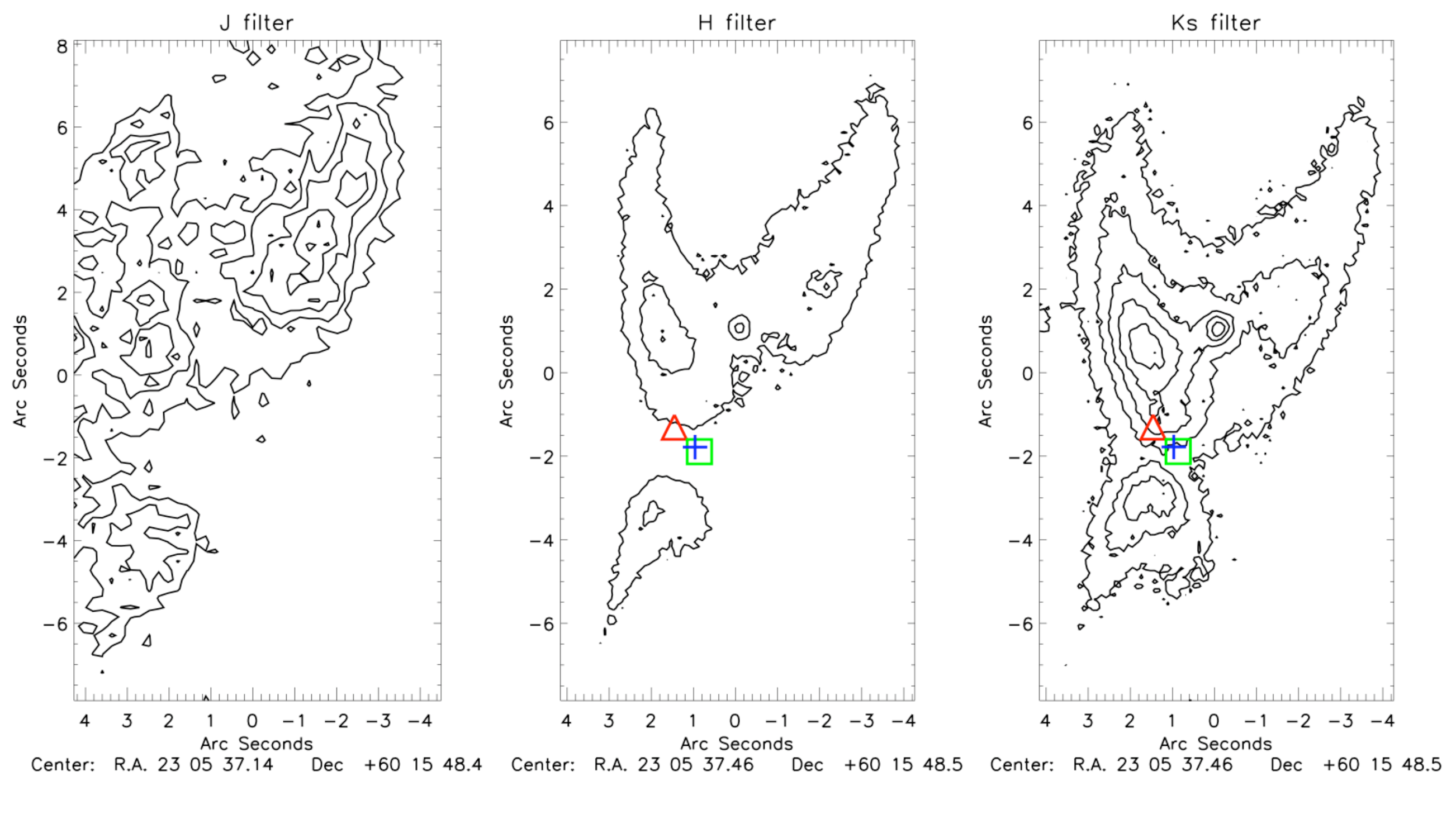}
\caption{Contours of the NIR images: $J$ filter (left), $H$ filter (middle), and $K'$ filter (right). In the $J$ band the S/N is lower than in the other filters and the 5 contour levels start at 5-$\sigma$ and go up to 13-$\sigma$ in equidistant steps. The lowest level corresponds to $\sim2.75\cdot10^{-4}$ mJy/pixel. In the $H$ and $K'$ filter images the S/N is higher and the flux density levels of the contours are identical for better comparison. Here, the lowest contour equals $\sim2.0\cdot10^{-4}$ mJy/pixel corresponding to $\sim$7-$\sigma$ in $H$ and $\sim$6-$\sigma$ in $K'$. The contours increase by $2.0\cdot10^{-4}$ mJy/pixel reaching thus $\sim6.0\cdot10^{-4}$ mJy/pixel in the third level (maximum level in the $H$ image) and $\sim1.0\cdot10^{-3}$ mJy/pixel in the fifth level (maximum level in the $K'$ image). The triangle, the square and the cross in the $H$ and $K'$ image denote the position of the flux peak of the IRAC 3.6 $\mu$m, 5.8 $\mu$m and SMA 1.3 mm emission, respectively.  The point-like structure seen in $H$ and $K'$ $\sim 1''$ north of the image center has very red colors and is presumably a background source shining through the nebula.
\label{nir-images}}
\end{figure}

\clearpage

\begin{figure}
\epsscale{2.}
\plotone{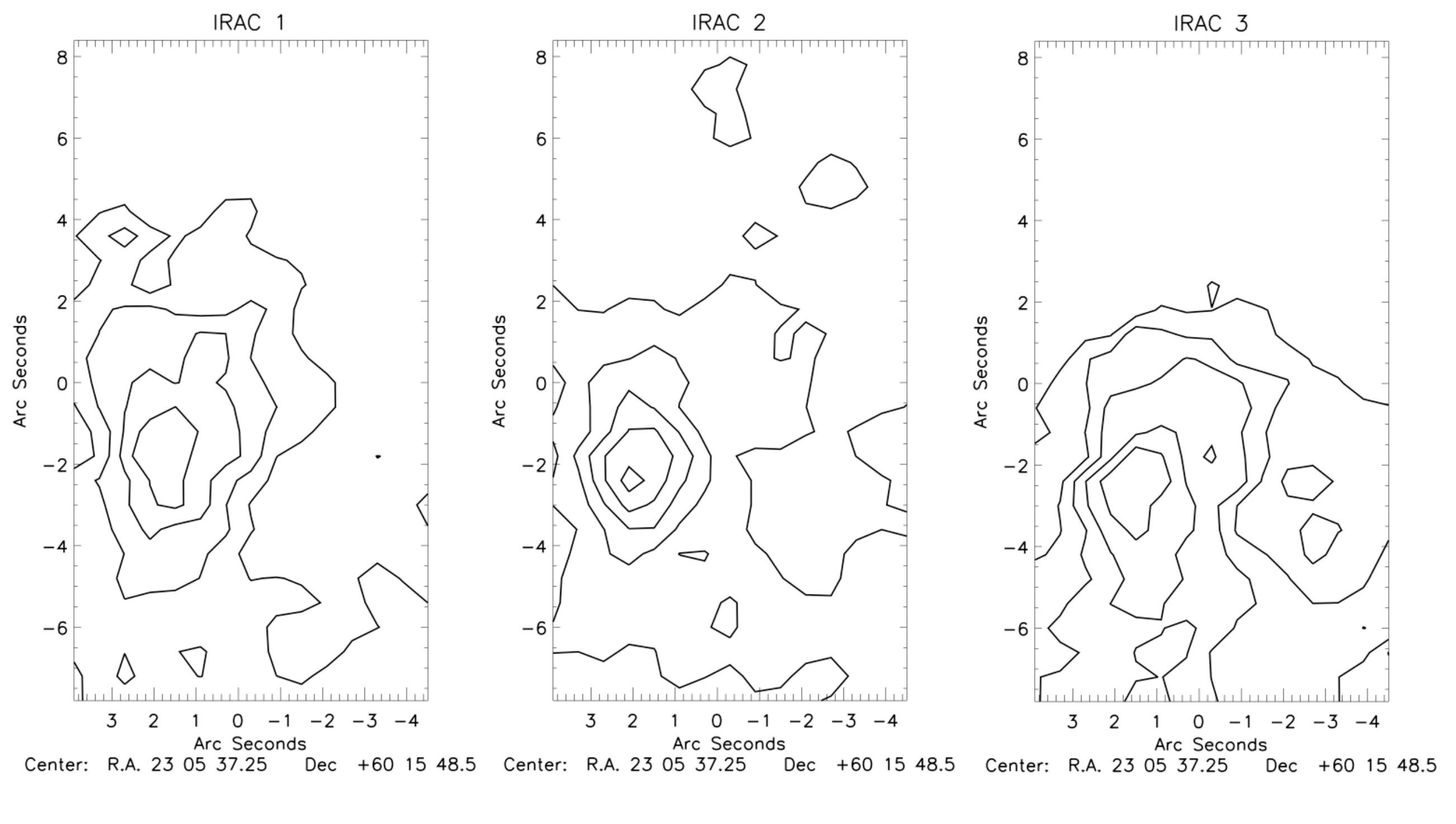}
\caption{Contours of the IRAC images: 3.6 $\mu$m (left), 4.5 $\mu$m (middle), and 5.8 $\mu$m (right). The lowest contour of the 3.6 $\mu$m image denotes a 6-$\sigma$ detection corresponding to $\sim$0.0144 mJy/pixel. The contour levels increase in 2-$\sigma$ steps yielding a 12-$\sigma$ level for the highest contour. For the 4.5 $\mu$m image the contours start also at the 6-$\sigma$ level corresponding to $\sim$0.0168 mJy/pixel but then increase in 3-$\sigma$ steps with a maximum of 18-$\sigma$. Finally, in the 5.8 $\mu$m image the emission rather strong possibly enhanced by diffuse PAH emission. The contours denote 33-$\sigma$ at the lowest level equaling $\sim$0.099 mJy/pixel and increase in 3-$\sigma$ steps to a maximum level of 45-$\sigma$. 
\label{irac-images}}
\end{figure}

\clearpage

\begin{figure}
\epsscale{2.}
\plotone{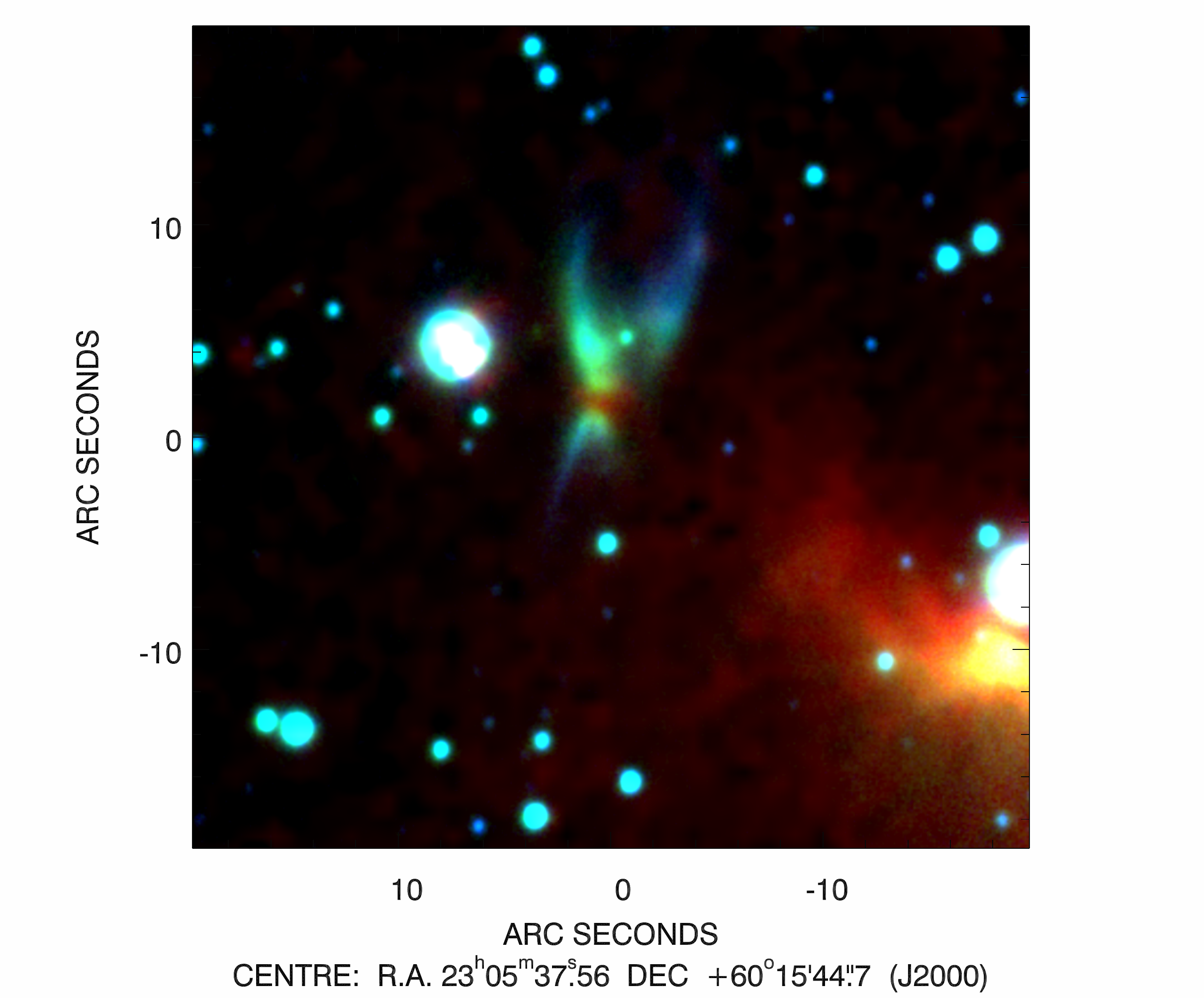}
\caption{Color composite of $H$ (blue), $K'$ (green) and IRAC 4.5 $\mu$m (red). While the bipolar nebula is only seen in scattered light in the $H$ and $K'$ band, the flux in the IRAC filter nicely traces the disk intersecting the nebula. 
\label{color-image-zoom}}
\end{figure}

\clearpage

\begin{figure}
\epsscale{1.}
\plotone{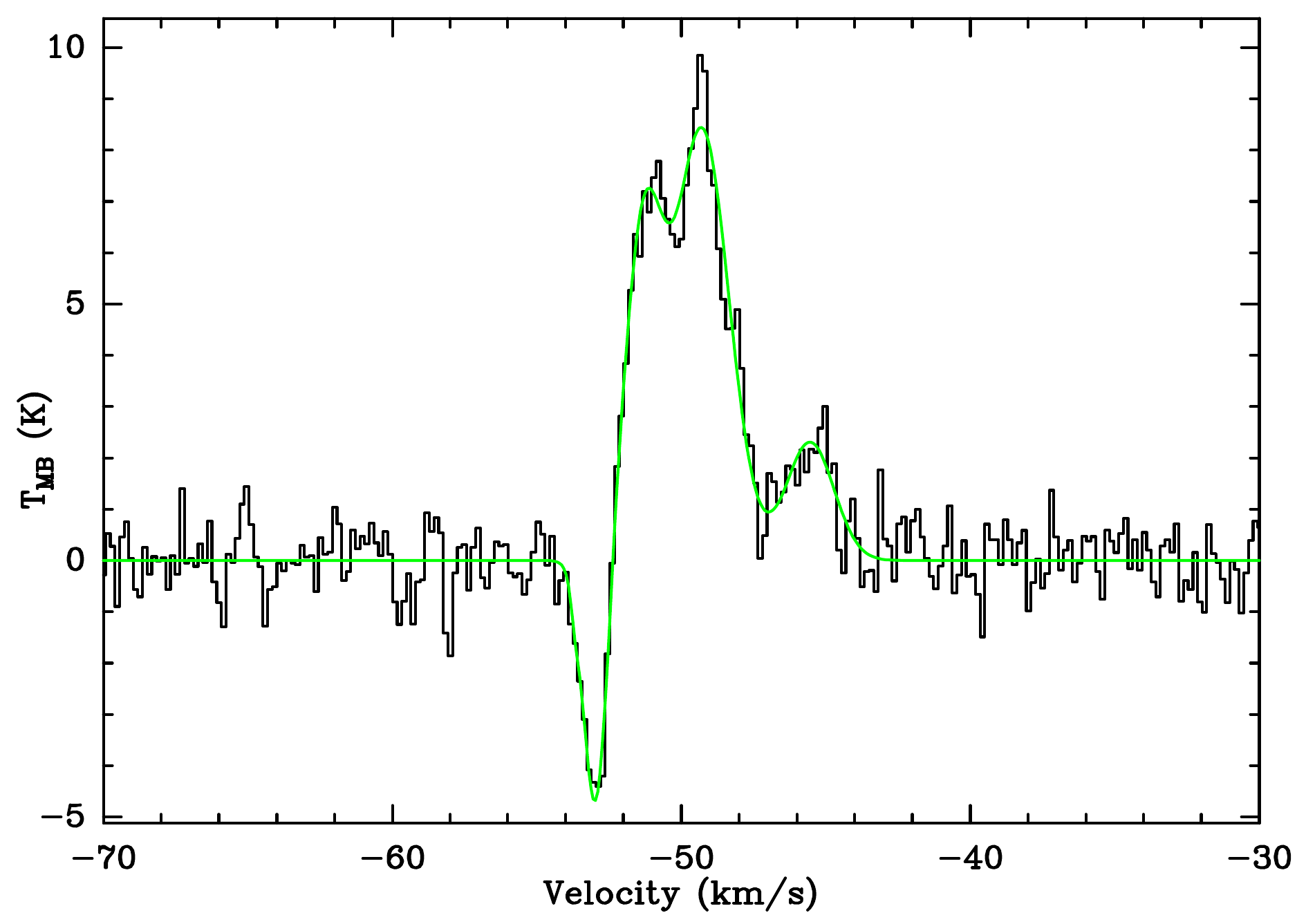}
\caption{Main beam brightness temperature of CO(3--2) transition line emission towards CAHA J23056+6016 measured with the 345 GHz receiver at the SMT. The parameter for gaussian fits to the different lines are given in Table~\ref{lines}.
\label{smt-spectrum}}
\end{figure}


\begin{figure}
\epsscale{.7}
\plotone{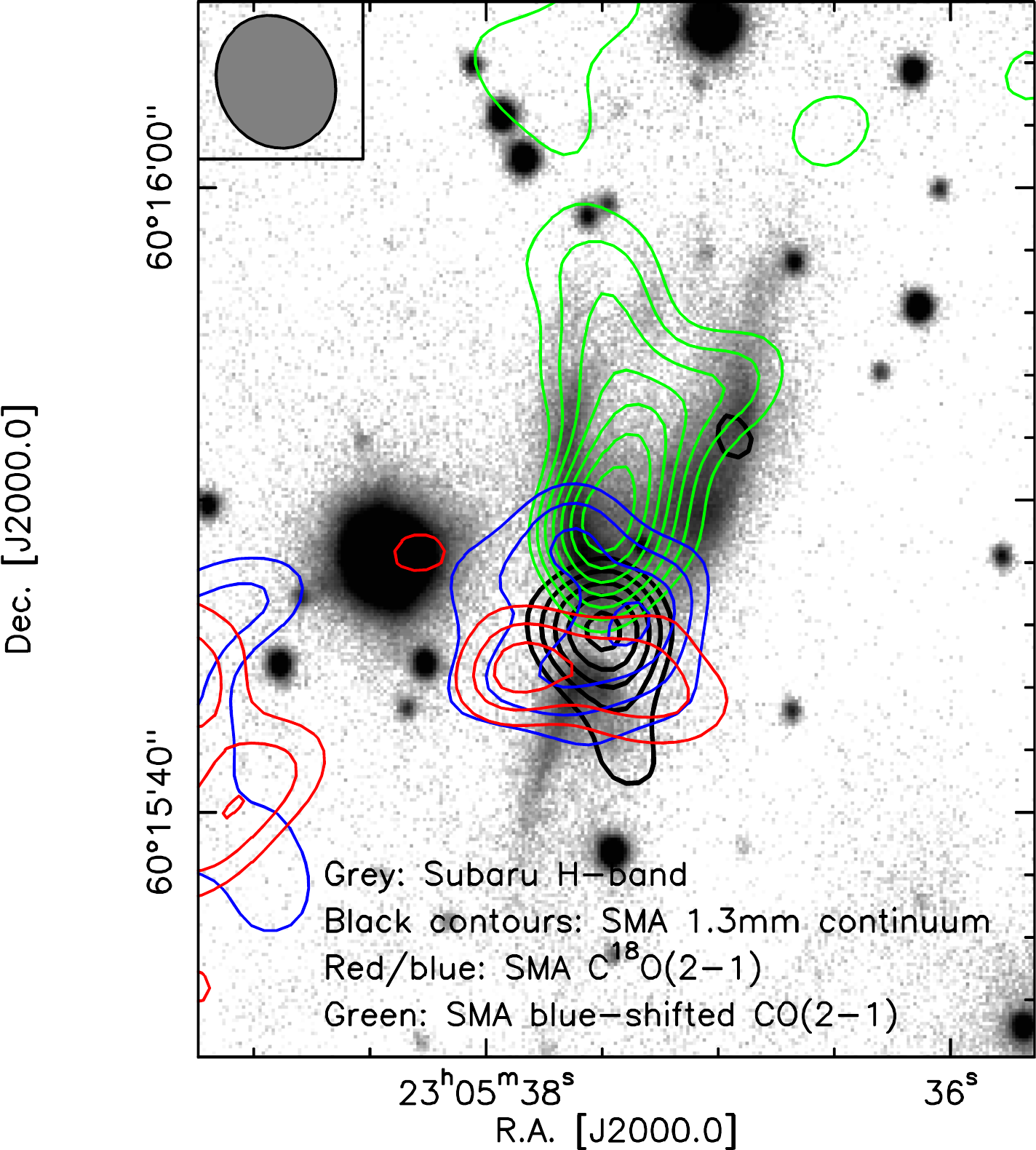}
\caption{Overlay of mm continuum and molecular line observations on the $H$-filter image. Black contours are 1.3 mm continuum emission, green contours blue-shifted CO(2--1), red/blue are C$^{18}$O(2--1). The continuum contours start at 3-$\sigma$ and increase in 1-$\sigma$ steps corresponding to 1.5 mJy/beam. The green contours show the integrated CO(2--1) emission between -58 and -54 km/s. They start at 3-$\sigma$ and increase in 1-$\sigma$ steps corresponding to 66 mJy/beam. The blue and red contours are centered on -51.6 and -51.0 km/s in velocity space, respectively, and each velocity channel has a width of 0.6 km/s. The lowest contour shows again the 3-$\sigma$ level and the contours increase in 1-$\sigma$ steps corresponding to 144 mJy/beam.
\label{mm-image}}
\end{figure}


\begin{figure}
\epsscale{1.}
\plotone{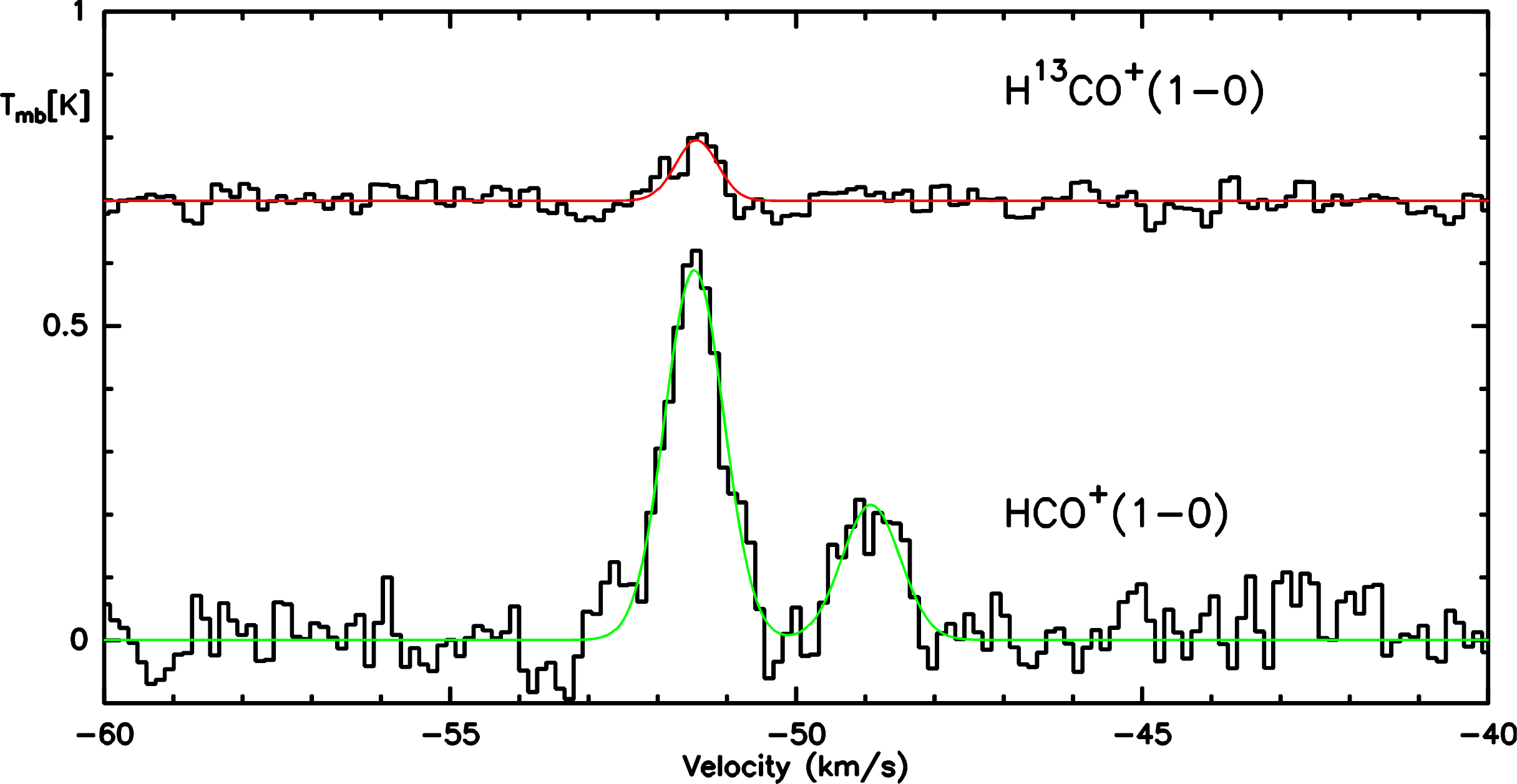}
\caption{Main beam brightness temperature of the HCO$^+$(1--0) and H$^{13}$CO$^+$(1--0) emission lines towards CAHA J23056+6016. The fit results (with set weight sigma) are summarized in Table~\ref{lines}.
\label{hco+-data}}
\end{figure}


\begin{figure}
\epsscale{1.}
\plotone{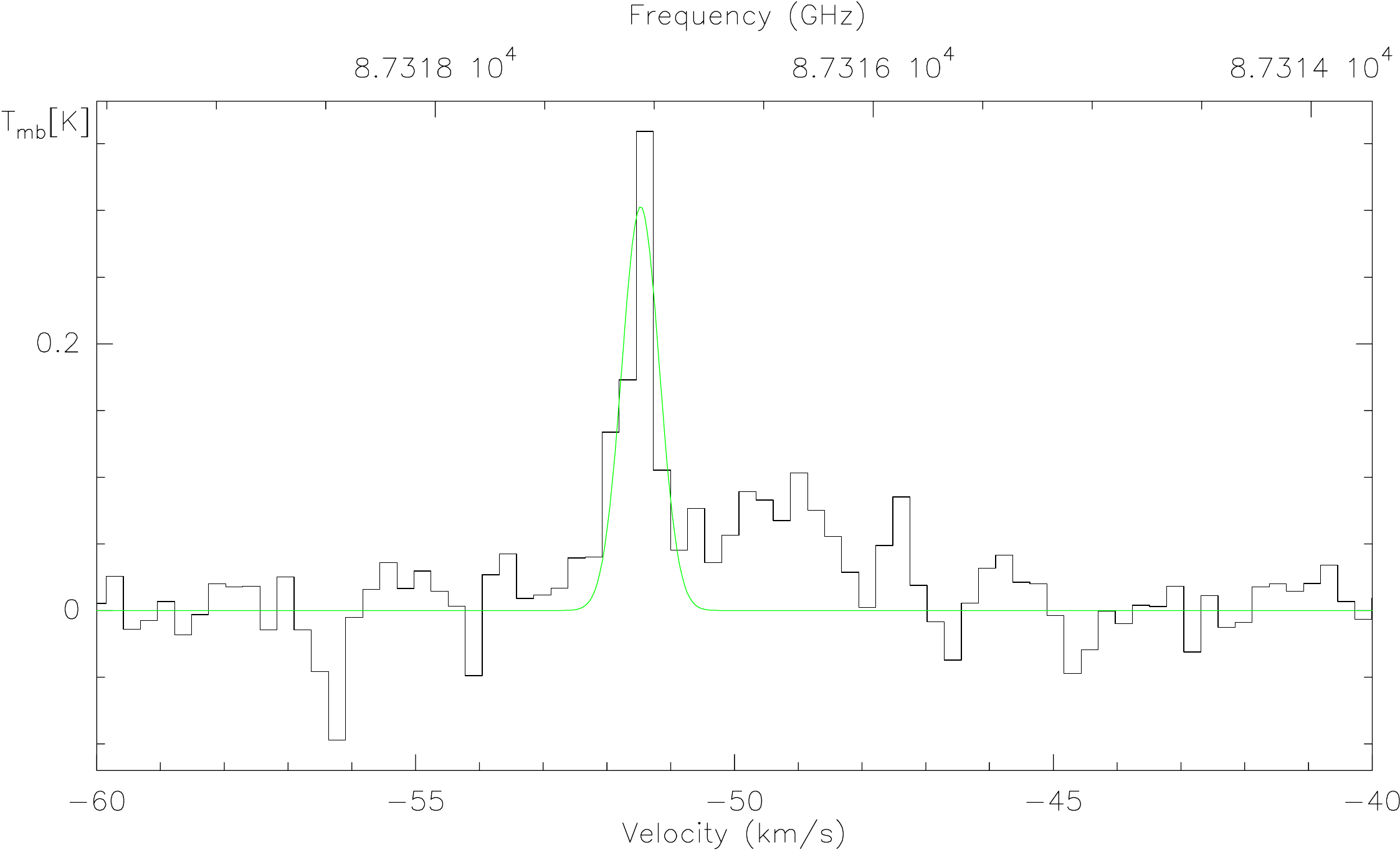}
\caption{Main beam brightness temperature of the C$_2$H(1--0) emission line towards CAHA J23056+6016. The fit parameter are summarized in Table~\ref{lines}.
\label{ch-data}}
\end{figure}

\clearpage

\begin{figure}
\epsscale{0.7}
\plotone{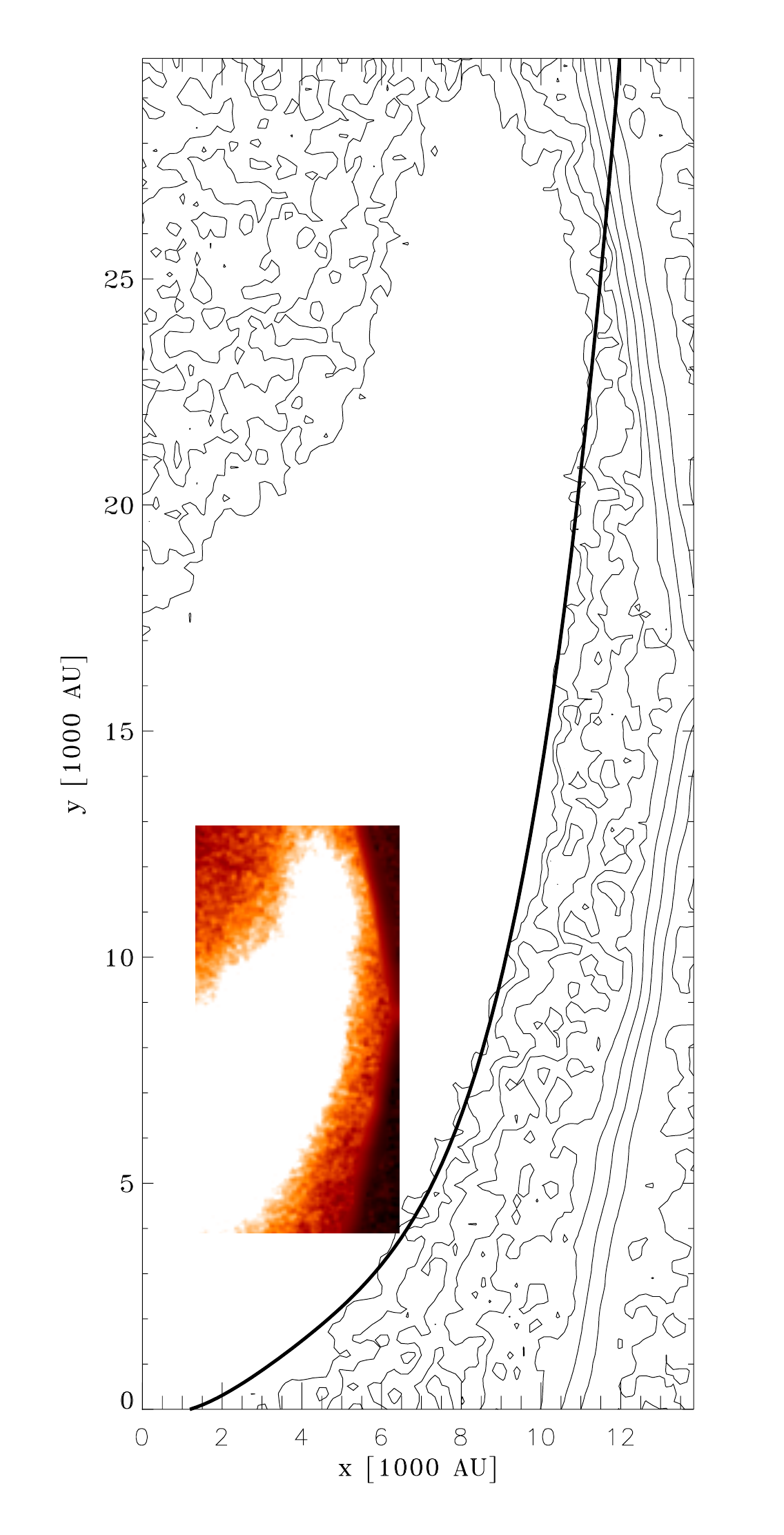}
\caption{Upper reflection nebula borders derived from a convolution of the $H$ and $K$
images assuming symmetry. The inlet shows the convolved image, the contour
plot contains the fit of the border by a polynomial as thick line.
\label{envelope_edge}}
\end{figure}


\begin{figure}
\epsscale{0.7}
\plotone{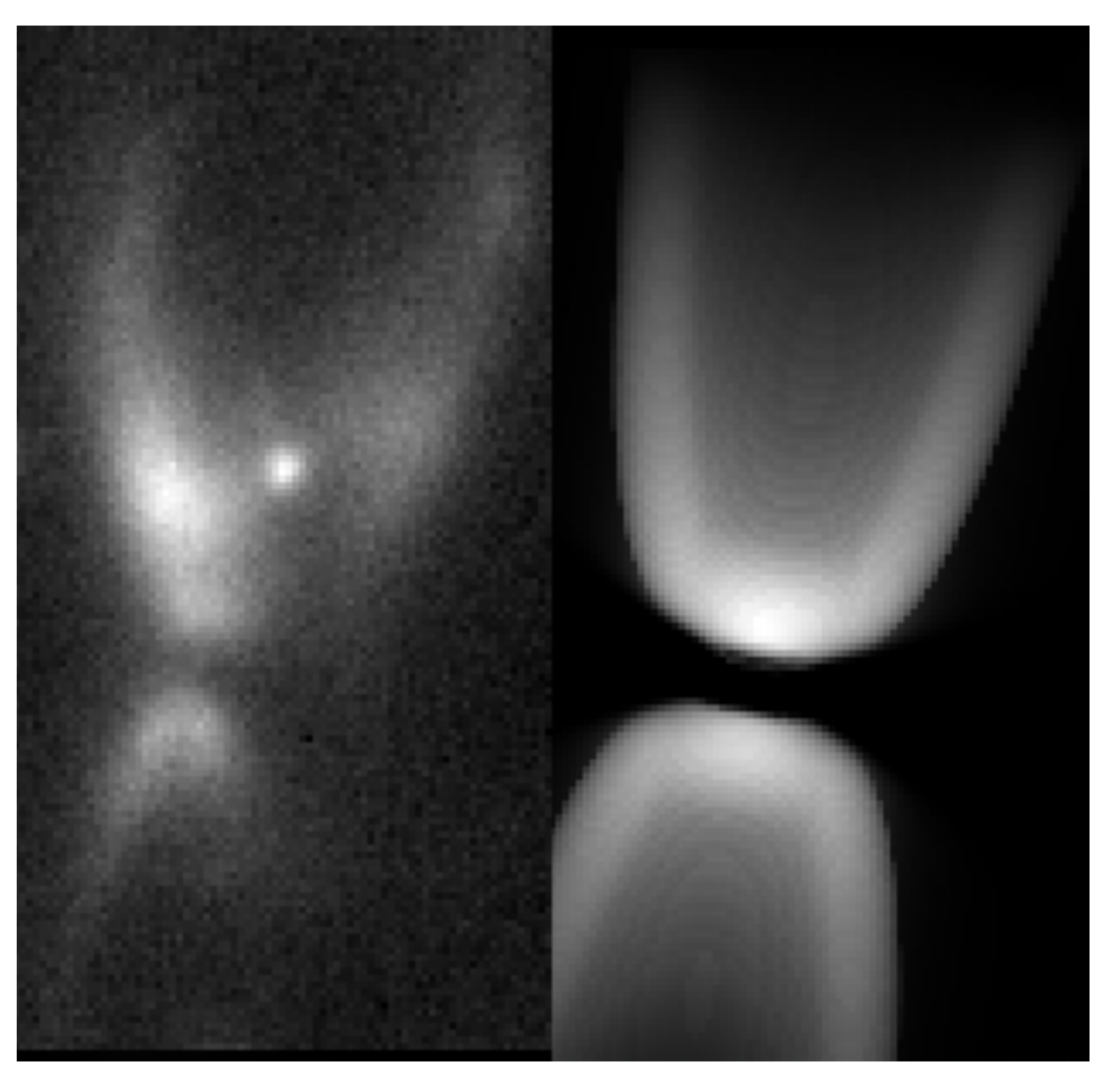}
\caption{$K$-band image (left) and scattered light
model image (right) of the reflection nebula
produced by a radiative transfer calculation based on a 2D disk plus
nebula density model.
\label{disk_model}}
\end{figure}

\clearpage

\end{document}